\documentclass[aps,pra,reprint,twocolumn,preprintnumbers,showpacs,superscriptaddress,amsmath,amssymb]{revtex4}
\usepackage{dcolumn}
\usepackage{bm}
\usepackage{amssymb}
\usepackage[german, english]{babel}
\usepackage{graphicx}
\usepackage{sidecap}
\usepackage{color}
\usepackage[letterpaper,total={7in,9.5in},top=0.75in,left=0.75in]{geometry}

\begin{document}
\title{Laser cooling to quantum degeneracy}

\author{Simon Stellmer}
\affiliation{Institut f\"ur Quantenoptik und Quanteninformation (IQOQI),
\"Osterreichische Akademie der Wissenschaften, 6020 Innsbruck,
Austria}
\author{Benjamin Pasquiou}
\affiliation{Institut f\"ur Quantenoptik und Quanteninformation (IQOQI),
\"Osterreichische Akademie der Wissenschaften, 6020 Innsbruck,
Austria}
\author{Rudolf Grimm}
\affiliation{Institut f\"ur Quantenoptik und Quanteninformation (IQOQI),
\"Osterreichische Akademie der Wissenschaften, 6020 Innsbruck,
Austria}
\affiliation{Institut f\"ur Experimentalphysik und
Zentrum f\"ur Quantenphysik, Universit\"at Innsbruck,
6020 Innsbruck, Austria}
\author{Florian Schreck}
\affiliation{Institut f\"ur Quantenoptik und Quanteninformation (IQOQI),
\"Osterreichische Akademie der Wissenschaften, 6020 Innsbruck,
Austria}

\date{\today}


\begin{abstract}
We report on Bose-Einstein condensation (BEC) in a gas of strontium atoms, using laser cooling as the only cooling mechanism. The condensate is formed within a sample that is continuously Doppler cooled to below 1\,$\mu$K on a narrow-linewidth transition. The critical phase-space density for BEC is reached in a central region of the sample, in which atoms are rendered transparent for laser cooling photons. The density in this region is enhanced by an additional dipole trap potential. Thermal equilibrium between the gas in this central region and the surrounding laser cooled part of the cloud is established by elastic collisions. Condensates of up to $10^5$ atoms can be repeatedly formed on a timescale of 100\,ms, with prospects for the generation of a continuous atom laser.
\end{abstract}

\maketitle

Laser cooling has revolutionized contemporary atomic and molecular physics in many respects, for example pushing the precision of clocks by orders of magnitude, and enabling ion quantum computation \cite{Arimondo1992book}. Since the early days of laser cooling, the question has been asked if the quantum degenerate regime could be reached using this efficient method as the only cooling process. Despite significant experimental and theoretical effort to overcome the limitations of laser cooling this goal has been elusive. So far, laser cooling had to be followed by evaporative cooling to reach quantum degeneracy \cite{Inguscio1999book}.

A gas of bosonic atoms with number density $n$ and temperature $T$ enters the quantum-degenerate regime and forms a Bose-Einstein condensate if its phase-space density $n \lambda_{dB}^3$ exceeds a critical value of 2.612. Here, $\lambda_{dB}=h/(2 \pi m k_B T)^{1/2}$ is the thermal de Broglie wavelength, where $h$ and $k_B$ are Planck's and Boltzmann's constant, respectively, and $m$ is the mass of an atom. Since $n \lambda_{dB}^3 \propto n T^{-3/2}$, low temperatures in combination with high densities have to be reached to obtain quantum degeneracy. Numerous studies, mainly carried out in the 1980's and 90's, have paved the ground to the present state of the art of laser cooling and have identified the limitations of this technique \cite{Chu1989lca}.

\begin{figure}
\includegraphics[width=86mm]{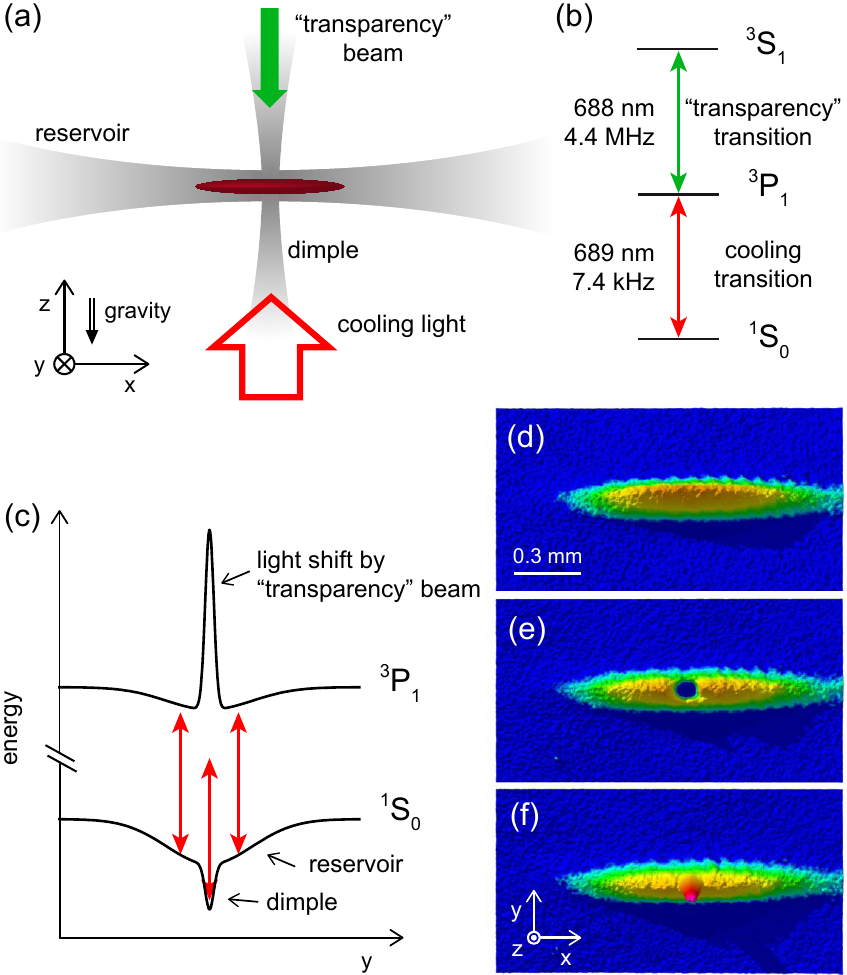}
\caption{Scheme to reach quantum degeneracy by laser cooling. (a) A cloud of atoms is confined in a deep reservoir dipole trap and exposed to a single laser cooling beam (red arrow). Atoms are rendered transparent by a ``transparency'' laser beam (green arrow) and accumulate in a dimple dipole trap by elastic collisions. (b) Level scheme showing the laser cooling transition and the transparency transition. (c) Potential experienced by $^1S_0$ ground-state atoms and atoms excited to the $^3P_1$ state. The transparency laser induces a light shift on the $^3P_1$ state, which tunes the atoms out of resonance with laser cooling photons. (d) to (f) Absorption images of the atomic cloud recorded using the laser cooling transition. The images show the cloud from above and demonstrate the effect of the transparency laser (e) and the dimple (f). (d) is a reference image without these two laser beams.}
\label{fig:Fig1}
\end{figure}

The longstanding goal of reaching the quantum degenerate regime by laser cooling \cite{Dum1994lct,Cirac1996clc,Morigi1998lco,Castin1998rol,Santos1999dco} can be discussed in terms of three main experimental challenges. First, temperatures in the low microkelvin regime have to be reached. Only here, quantum degeneracy can be obtained at a density that is low enough to avoid fast decay of the gas by molecule formation. This challenge has been met with several laser cooling techniques, as for example Sisyphus cooling \cite{Lett1988ooa,Dalibard1989lcb}, velocity selective coherent population trapping \cite{Aspect1988lcb}, Raman cooling \cite{Kasevich1992lcb}, Raman sideband cooling \cite{Hamann1998rsr}, or Doppler cooling on narrow lines \cite{Ido2000odt}. The second challenge is the implementation of an efficient trapping scheme that allows for accumulation of atoms at high density in a particular region \cite{StamperKurn1998rfo,Weber2003bec}. The third, and most severe challenge is to avoid the detrimental effects of the laser cooling photons, which impede the required density increase. One such effect are light-assisted inelastic collisions, which lead to loss \cite{Julienne1992toc,Walker1997moc}. Another is the reabsorption of photons scattered during laser cooling \cite{Sesko1991bon}, which leads to an effective repulsion between the atoms and to heating of atoms in the lowest energy states. Both effects increase with density and make it impossible to reach quantum degeneracy. For low phase-space density samples, this challenge has been overcome by rendering the atoms transparent to laser cooling photons \cite{Ketterle1993hdo,Griffin2006ssl,Clement2009aor} or by decreasing the photon scattering rate below the frequency of a confining trap \cite{Cirac1996clc,Castin1998rol,Wolf2000sor}. It has also been proposed to reduce reabsorption by dimensional reduction of the sample \cite{Castin1998rol}. The solutions to the three challenges implemented so far are insufficient to reach quantum degeneracy. The highest phase-space densities ever attained are one order of magnitude too low \cite{Han20003DR,Ido2000odt}. Surprisingly, this last order of magnitude has been an insurmountable obstacle for a decade.

In this Letter, we present an experiment hat overcomes all three challenges and creates a BEC of strontium by laser cooling. Our scheme is based on the combination of three techniques, favored by the properties of this element, and does not rely on evaporation. Strontium possesses a transition with such a narrow linewidth ($\Gamma/2\pi=7.4\,$kHz) that simple Doppler cooling can reach temperatures down to 350\,nK \cite{Ido2000odt,Loftus2004nlc,Boyd2007Thesis}. Using this transition, we prepare a laser cooled sample of $10^7$ $^{84}$Sr atoms in a large ``reservoir'' dipole trap. To avoid the detrimental effects of laser cooling photons, we render atoms transparent for these photons in a small spatial region within the laser cooled cloud. Transparency is induced by a light shift on the optically excited state of the laser cooling transition. In the region of transparency, we are able to increase the density of the gas, by accumulating atoms in an additional, small ``dimple'' dipole trap \cite{StamperKurn1998rfo,Weber2003bec}. Atoms in the dimple thermalize with the reservoir of laser-cooled atoms by elastic collisions and form a BEC.

The details of our scheme are shown in Fig.~\ref{fig:Fig1}. Based on our previous work \cite{Stellmer2009bec,Stellmer2012pod,Stellmer2013Thesis}, we use several stages of laser cooling to prepare a sample of $^{84}$Sr atoms in the reservoir trap \cite{EndnoteMaterialsAndMethods}. The trap consists of an infrared laser beam (wavelength 1065\,nm) propagating horizontally ($x$-direction). The beam profile is strongly elliptic, with a beam waist of 300\,$\mu$m in the horizontal direction ($y$-direction) and 17\,$\mu$m along the field of gravity ($z$-direction). The depth of the reservoir trap is kept constant at $k_B \times 9\,\mu$K. After preparation of the sample, another laser cooling stage is performed on the narrow $^1S_0 - {^3P_1}$ intercombination line, using a single laser beam propagating vertically upwards. The detuning of the laser cooling beam from resonance is $\sim -2\,\Gamma$ and the peak intensity is $0.15\,\mu$W/cm$^2$, which is 0.05 of the transition's saturation intensity. These parameters result in a photon scattering rate of $\sim 70\,$s$^{-1}$. At this point, the ultracold gas contains $9\times 10^6$ atoms at a temperature of 900\,nK.

To render the atoms transparent to cooling light in a central region of the laser cooled cloud, we induce a light shift on the $^3P_1$ state, using a ``transparency'' laser beam 15\,GHz blue detuned to the $^3P_1 - {^3S_1}$ transition \cite{EndnoteMaterialsAndMethods}. This beam propagates downwards under a small angle of $15^{\circ}$ to vertical and has a beam waist of 26\,$\mu$m in the plane of the reservoir trap ($xy$-plane). The beam has a peak intensity of $0.7\,$kW/cm$^2$. It upshifts the $^3P_1$ state by more than 10\,MHz and also influences the nearest molecular level tied to the $^3P_1$ state significantly \cite{Stellmer2012cou,EndnoteMaterialsAndMethods}. Related schemes of light-shift engineering were used to image the density distribution of atoms \cite{Thomas1995ppm,Brantut2008lst}, to improve spectroscopy \cite{Kaplan2002soi}, or to enhance loading of dipole traps \cite{Griffin2006ssl,Clement2009aor}. To demonstrate the effect of the transparency laser beam, we take absorption images of the cloud on the laser cooling transition. Figure~\ref{fig:Fig1}(d) shows a reference image without the transparency beam. In presence of this laser beam, atoms in the central part of the cloud are transparent for the probe beam, as can be seen in Fig.~\ref{fig:Fig1}(e).

To increase the density of the cloud, a dimple trap is added to the system. It consists of an infrared laser beam (wavelength 1065\,nm) propagating upwards under a small angle of $22^{\circ}$ to vertical and crossing the laser cooled cloud in the region of transparency. In the plane of the reservoir trap, the dimple beam has a waist of 22\,$\mu$m. The dimple is ramped to a depth of $k_B \times 2.6\,\mu$K, where it has trap oscillation frequencies of 250\,Hz in the horizontal plane. Confinement in the vertical direction is only provided by the reservoir trap and results in a vertical trap oscillation frequency of 600\,Hz. Figure~\ref{fig:Fig1}(f) shows a demonstration of the dimple trap in absence of the transparency beam. The density in the region of the dimple increases substantially. However, with the dimple alone no BEC is formed because of photon reabsorption.

\begin{figure*}
\includegraphics[width=179mm]{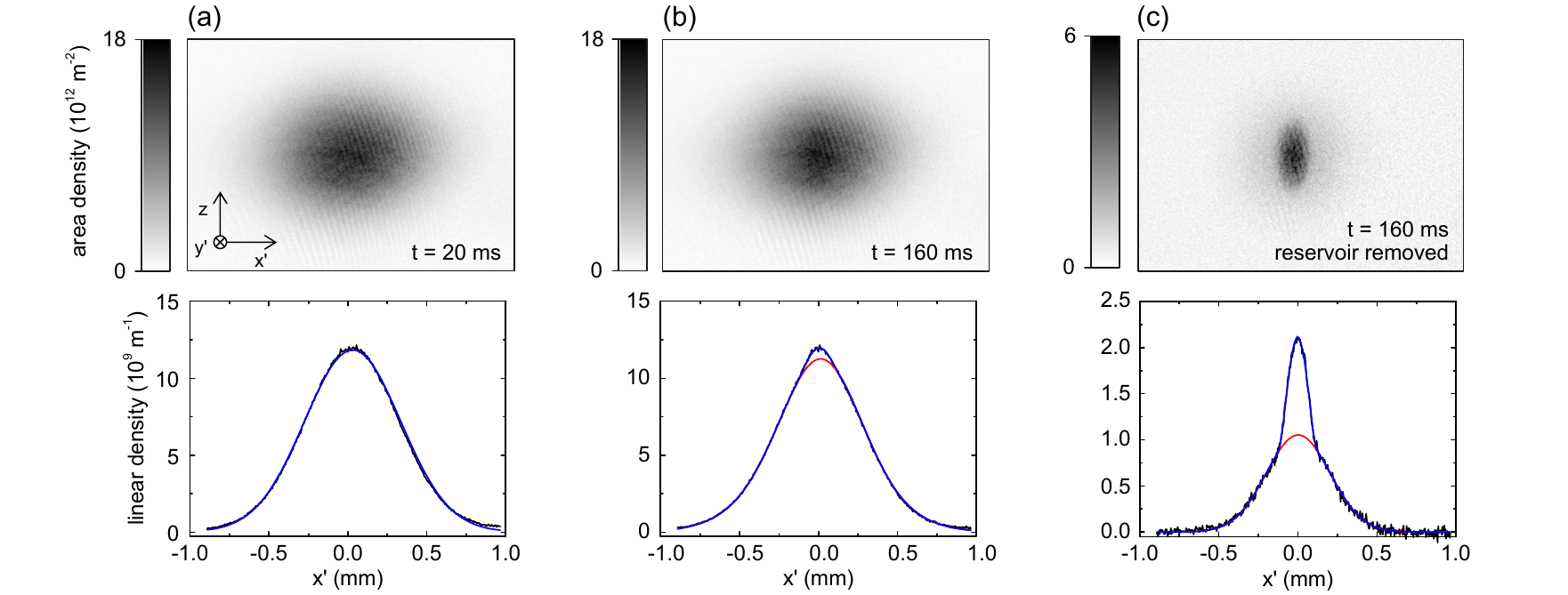}
\caption{\label{fig:Fig2} Creation of a BEC by laser cooling. Shown are time-of-flight absorption images and integrated density profiles of the atomic cloud for different times $t$ after the transparency laser has been switched on, recorded after 24\,ms of free expansion. (a) and (b) The appearance of an elliptic core at $t=160\,$ms indicates the creation of a BEC. (c) Same as in (b), but to increase the visibility of the BEC, atoms in the reservoir trap were removed before the image was taken. The fits (blue lines) consist of Gaussian distributions to describe the thermal background and an integrated Thomas-Fermi distribution describing the BEC. The red lines show the component of the fit corresponding to the thermal background. The $x^\prime y^\prime$-plane is rotated by $45^\circ$ around the $z$-axis with respect to the $xy$-plane and the field of view of the absorption images is 2\,mm $\times$ 1.4\,mm.}
  \label{fig:Fig2}
\end{figure*}

The combination of the transparency laser beam and the dimple trap leads to BEC. Starting from the laser cooled cloud held in the reservoir trap, we switch on the transparency laser beam and ramp the dimple trap to a depth of $k_B \times 2.6\,\mu$K. The potentials of the $^1S_0$ and $^3P_1$ states in this situation are shown in Fig.~\ref{fig:Fig1}(c). Atoms accumulate in the dimple without being disturbed by photon scattering. Elastic collisions thermalize atoms in the dimple with the laser cooled reservoir. The phase-space density in the dimple increases and a BEC emerges.

We detect the BEC by taking absorption images 24\,ms after switching off all laser beams. Figure~\ref{fig:Fig2}(a) shows the momentum distribution 20\,ms after switching on the transparency beam, which is well described by a thermal distribution. By contrast, we observe that 140\,ms later, an additional, central elliptical feature has developed; see Fig.~\ref{fig:Fig2}(b). This is the hallmark of the BEC. Although clearly present, the BEC is not very well visible in Fig.~\ref{fig:Fig2}(b), because it is shrouded by $8\times 10^6$ thermal atoms originating from the reservoir. To show the BEC with higher contrast, we have developed a background reduction technique. We remove the reservoir atoms by an intense flash of light on the $^1S_0-^3P_1$ transition applied for 10\,ms. Atoms in the region of transparency remain unaffected by this flash. Only $5\times10^5$ thermal atoms in the dimple remain and the BEC stands out clearly; see Fig.~\ref{fig:Fig2}(c). We use this background reduction technique only for demonstration purposes, but not for measuring atom numbers or temperatures.

Quantitative data on our experiment are obtained by two-dimensional fits to time-of-flight absorption images \cite{EndnoteMaterialsAndMethods}. The atom number of the thermal cloud and of the BEC are extracted from fits to 24-ms expansion images, consisting of Gaussian distributions describing the thermal background and an integrated Thomas-Fermi distribution describing the BEC. Further absorption images taken after 4\,ms expansion time are used to determine atom number and temperature of the gas in the reservoir and the dimple, respectively.

\begin{figure}
\includegraphics[width=86mm]{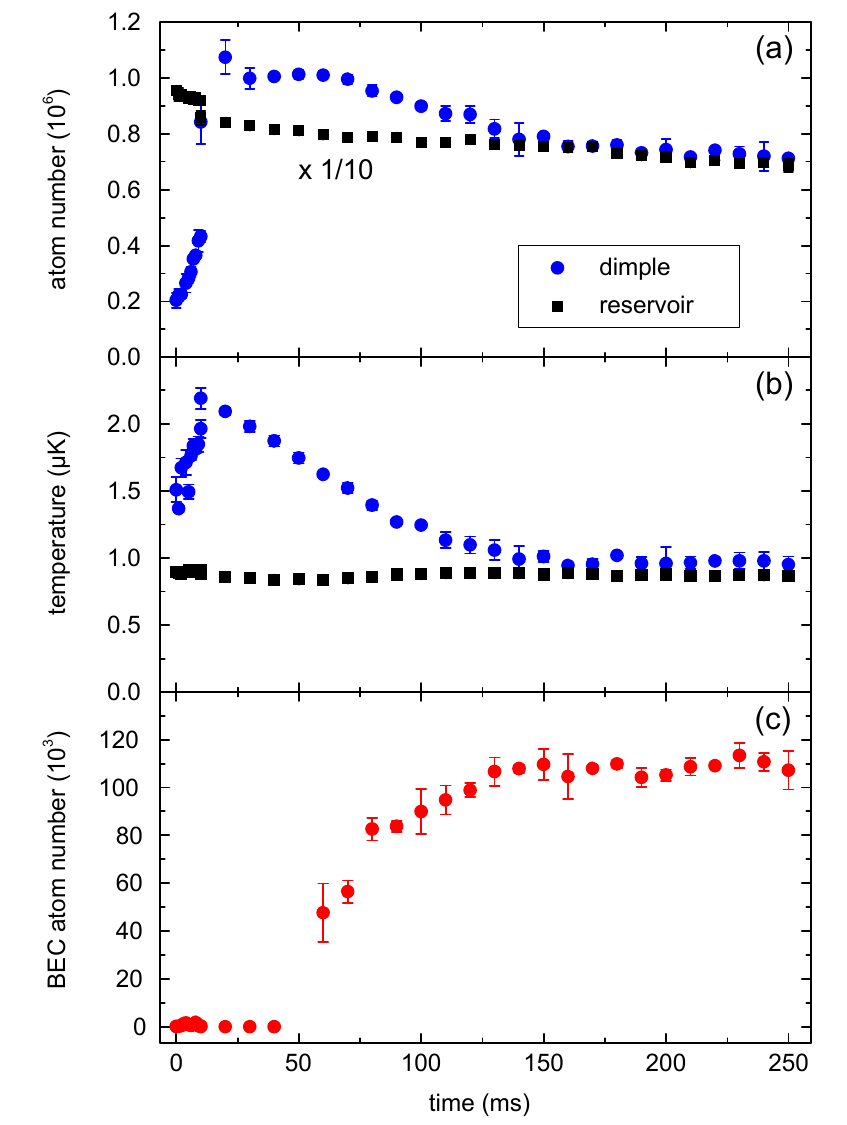}
\caption{\label{fig:Fig3} Characterization of the BEC formation process after the transparency laser is switched on. Shown are the evolution of the atom number in the dimple and the reservoir (a), the evolution of temperature in these regions (b) and the BEC atom number (c). During the first 10\,ms of this evolution, the dimple trap is ramped on. After 60\,ms a BEC is detected.}
\end{figure}

We now analyze the dynamics of the system after the transparency laser beam has been switched on. As we increase the dimple strength to its final depth in 10\,ms, $10^6$ atoms accumulate in it and the temperature of the dimple gas increases; see Figs.~\ref{fig:Fig3}(a) and (b). During the next $\sim 100\,$ms the dimple gas thermalizes with the reservoir gas by elastic collisions \cite{Kohl2002gob,Ritter2007otf}. The temperature of the reservoir gas is hereby not increased, since the energy transferred to it is dissipated by laser cooling. We carefully check that evaporation is negligible even for the highest temperatures of the gas \cite{EndnoteMaterialsAndMethods}. Already after 60\,ms a BEC is detected. Its atom number saturates at $1.1\times 10^5$ after 150\,ms, as shown in Fig.~\ref{fig:Fig3}(c). The atom number in the reservoir decreases slightly, initially because of migration into the dimple and on longer timescales because of light assisted loss processes in the laser cooled cloud.

\begin{figure}[t]
\includegraphics[width=86mm]{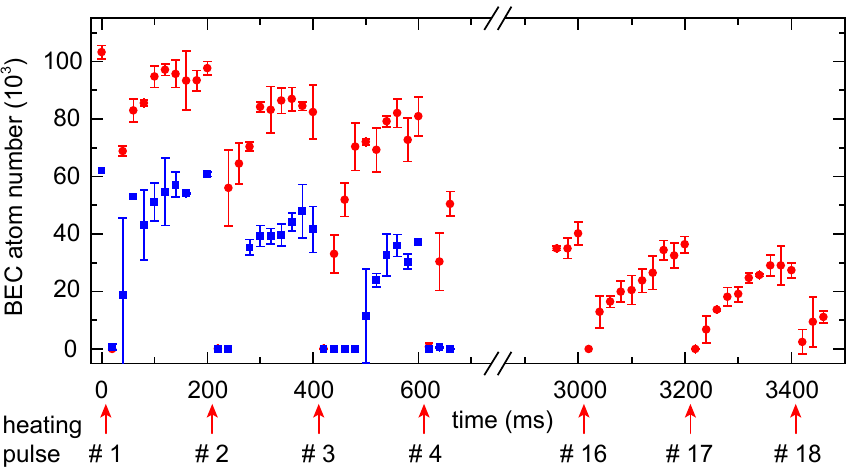}
\caption{\label{fig:Fig4} Repeated destruction and reformation of the BEC. Shown is the evolution of BEC atom number while the BEC is destroyed every 200\,ms (arrows) by suddenly increasing the depth of the dimple trap. If the system is laser cooled, the BEC atom number quickly increases again, which is shown here for up to 18 cycles (red circles). Without laser cooling, a BEC is detectable for at most five heating cycles, of which the first three are shown here (blue squares).}
\end{figure}

To demonstrate the power of our laser cooling scheme, we repeatedly destroy the BEC and let it reform (Fig.~\ref{fig:Fig4}). To destroy the BEC, we pulse the dimple trap depth to $k_B \times 15\,\mu$K for 2\,ms, which increases the temperature of the dimple gas by a factor two. We follow the evolution of the BEC atom number while the heating pulse is applied every 200\,ms. A new BEC starts forming a few 10\,ms after each heating pulse for more than 30 pulses. We find that the observed decrease in the BEC size from pulse to pulse stems from the reduction of the total atom number in the system.

To clarify the role laser cooling plays in our scheme, we perform a variation of the experiment. Here, we switch off the laser cooling beam before ramping up the dimple and we do not use the transparency beam. Heat released while ramping up the dimple or after a heating pulse is again distributed from the dimple to the whole system by elastic collisions, but this time not dissipated by laser cooling. Since the reservoir gas has a ten times higher atom number than the dimple gas, the temperature after thermalization is only increased by a small amount. If the final temperature in the dimple is below the critical temperature, a BEC is formed. This scheme resembles the formation of a BEC by trap deformation, as demonstrated in \cite{StamperKurn1998rfo} using a sample of atoms cooled by evaporation. We test the performance of this BEC creation scheme again by repeated heating pulses. We can detect a BEC after at most five pulses. For more pulses, the temperature of the gas in the dimple remains too high to allow the formation of a BEC. This poor behavior stands in stark contrast to the resilience of BEC formation to heating, if the system is continuously laser cooled.

The ability to reach the quantum degenerate regime by laser cooling has many exciting prospects. This method can be applied to any element possessing a laser cooling transition with a linewidth in the kHz range and suitable collision properties. Besides strontium this encompasses several lanthanides \cite{Lu2011sdb,Aikawa2012bec}. The technique can also cool fermions to quantum degeneracy and it can be extended to sympathetic cooling in mixtures of isotopes or elements. Another tantalizing prospect enabled by variations of our techniques, is the realization of a continuous atom laser, which converts a thermal beam into a laser-like beam of atoms. To realize such a device, our scheme needs to be extended in two ways. First, the reservoir of laser cooled atoms needs to be replenished, for example by sending a thermal beam of atoms onto a part of the reservoir with sufficiently high cooling laser intensity to allow capture of these atoms. Second, a continuous beam of condensed atoms needs to be outcoupled. Using magnetic species such as dysprosium or erbium, outcoupling from the BEC is possible by changing the internal state and thereby the magnetic force on the atoms \cite{Mewes1997ocf,Bloch1999alw}. Alternatively, the reservoir can be connected to an outcoupling dipole trap, creating a narrow channel where atoms can escape without further interaction with the cooling light \cite{Lahaye2005eco}.\\

We thank Matteo Zaccanti for careful reading of the manuscript. We gratefully acknowledge support from the Austrian Ministry of Science and Research (BMWF) and the Austrian Science Fund (FWF) through a START grant under Project No.~Y507-N20. As member of the project iSense, we also acknowledge financial support of the Future and Emerging Technologies (FET) program within the Seventh Framework Programme for Research of the European Commission, under FET-Open grant No.~250072.

\newpage

\setcounter{figure}{0}
\setcounter{table}{0}
\setcounter{equation}{0}
\renewcommand{\thefigure}{S\arabic{figure}}
\renewcommand{\thetable}{S\Roman{table}}
\renewcommand{\theequation}{S\arabic{equation}}

\begin{center}{\large{\textbf{Supplemental Material}}} \end{center}
\vspace{2mm}

This Supplemental Material contains in-depth information about our approach to reach quantum degeneracy by laser cooling. In Sec.~\ref{sec:Sequence} we give details on the experimental sequence. In Sec.~\ref{sec:Characterization} we discuss the parameter dependence of BEC creation and properties of the BEC. In Sec.~\ref{sec:Evaporation} we show that evaporative cooling does not play a role in reaching quantum degeneracy. In Sec.~\ref{sec:ESD} we analyze the shifting of atomic and molecular transitions by the transparency beam. In Sec.~\ref{sec:Model} we model the density distributions of the BEC and the thermal cloud.

\section{Experimental sequence}
\label{sec:Sequence}

In the following, we discuss in detail the experimental sequence with which we obtain BEC by laser cooling. We first describe the preparation of a sample of ultracold $^{84}$Sr atoms in the reservoir dipole trap, which follows closely the procedure used in our previous work \cite{Stellmer2009bec,Stellmer2012pod,Stellmer2013Thesis}. Then we give details on the additional steps we take to produce a BEC. These details include information on the transparency beam, the dimple dipole trap and the conditions of the final laser cooling stage. We end with a discussion of our data acquisition and analysis method.

\subsection{Sample preparation}

\textbf{Isotope choice ---} Of the three bosonic strontium isotopes, $^{84}$Sr is best suited for our experiment, since its scattering length of $a_{84}=124\,a_0$ (with $a_0$ the Bohr radius) allows efficient thermalization by elastic collisions. The other bosonic isotopes are unsuitable for our task. $^{88}$Sr has a negligible scattering length of $a_{88}=-1.4\,a_0$ and therefore does not thermalize. $^{86}$Sr suffers from three-body inelastic loss because of a very large scattering length $a_{86}=800\,a_0$. Unfortunately $^{84}$Sr has a low natural abundance of only 0.56\%. To obtain a large sample, we accumulate metastable state atoms in a magnetic trap, as described below.

\begin{figure}[b]
\includegraphics[width=86mm]{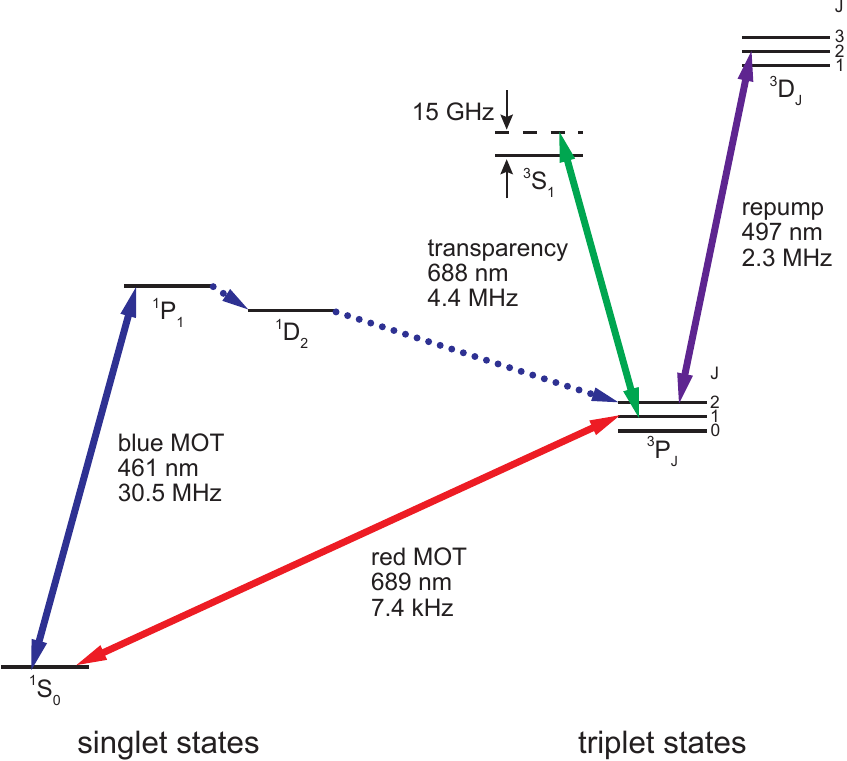}
\caption{Level scheme of strontium with all relevant states and transitions. The dotted arrows show the decay path of atoms from the $^1P_1$ state into the $^3P_2$ state.}
\label{fig:SrLevelScheme}
\end{figure}

\textbf{Atomic beam and blue MOT ---} A sample of strontium metal with natural isotopic composition is heated in an oven under vacuum to about $600\,^{\circ}$C in order to sublimate strontium atoms. The atoms form an atomic beam after escaping the oven through a bundle of microtubes. $^{84}$Sr atoms in the atomic beam are transversally cooled, Zeeman-slowed, and captured in a ``blue'' magneto-optical trap (MOT) with laser light red-detuned to the $^1S_0 - {^1P_1}$ transition at 461\,nm; see Fig.~\ref{fig:SrLevelScheme}. This transition has a linewidth of $\Gamma_{\rm blue}/2\pi=30.5\,$MHz, corresponding to a Doppler temperature of $T_{D, {\rm blue}}=\hbar \Gamma_{\rm blue} / (2 k_B)=720\,\mu$K. The MOT uses a quadrupole magnetic field with vertically oriented axis and a vertical field gradient of 55\,G/cm. The MOT beams have waists of 5\,mm and peak intensities of 10\,mW/cm$^2$, which is a quarter of the saturation intensity $I_{\mathrm{sat, blue}}=\pi h c \Gamma_{\rm blue}/3\lambda_{\rm blue}^3=40.7\,$mW/cm$^2$. The detuning of the MOT light to resonance is $-30$\,MHz.

\textbf{Metastable state reservoir ---} The blue MOT cycle is not completely closed, as atoms in the excited $^1P_1$ state can decay into the metastable and magnetic $^3P_2$ state with a branching ratio of 1:150\,000. Atoms in this state can be trapped in the quadrupole magnetic field of the MOT. Since the lifetime of magnetically trapped $^3P_2$-state atoms of $\sim 30\,$s is about three orders of magnitude longer than the leak time scale of the blue MOT, metastable state atoms accumulate in the magnetic trap. We operate the MOT until about $10^8$ atoms are accumulated in this ``metastable state reservoir'', which takes $\sim 10\,$s. This accumulation of atoms allows us to use the $^{84}$Sr isotope, despite its low natural abundance.

\textbf{Red MOT ---} To increase the phase-space density of the sample, we use a narrow-line ``red'' MOT operated on the $^1S_0 - {^3P_1}$ intercombination line, which has a wavelength of $\lambda_{\rm red}=689\,$nm and a linewidth of $\Gamma_{\rm red}/2\pi=7.4\,$kHz. Our red MOT laser has a linewidth of about 2\,kHz and an absolute stability better than 500\,Hz. The horizontal (upward, downward propagating) red MOT beams have a waist of 3.1\,mm (2.9\,mm, 3.1\,mm). The Doppler temperature of the red MOT $T_{D, {\rm red}}=180$\,nK is comparable to the recoil temperature $T_r=\hbar^2 k^2/(k_B m)=460\,$nK, where $k=2\pi/\lambda_{\rm red}$ is the wave vector of the cooling light, and $m$ is the atom's mass. The temperature can approach $T_r/2$ when reducing the cooling light intensity to the saturation intensity $I_{\mathrm{sat,red}}=3\,\mu$W/cm$^2$ \cite{Loftus2004nlc}. Not only the temperature but also the atom number decreases, when decreasing the cooling light intensity, and a compromise between atom number and temperature has to be chosen. We typically reach temperatures of 800\,nK with samples of about $10^7$\,atoms.

\begin{figure*}
\includegraphics[width=177.5mm]{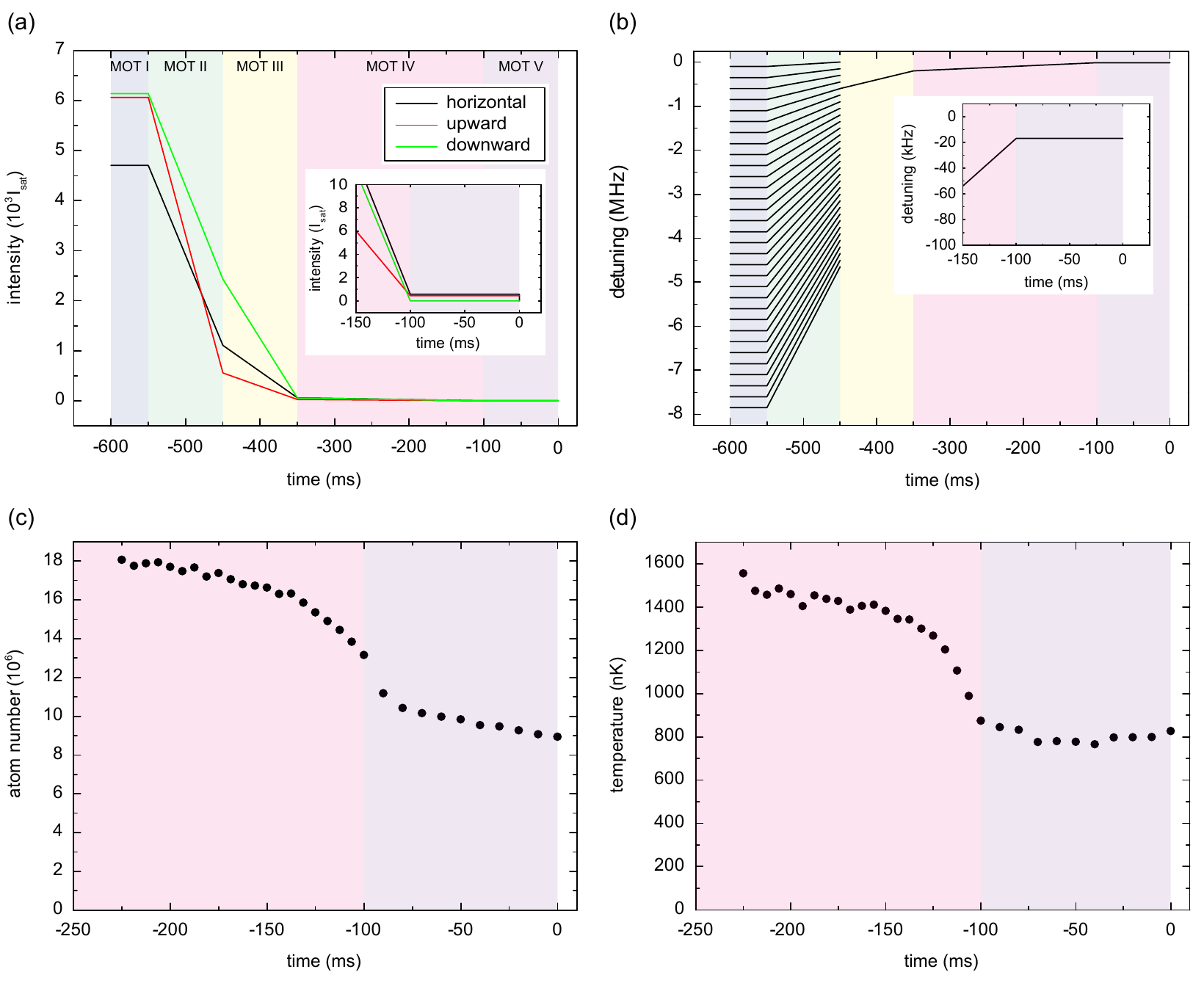}
\caption{\label{fig:red_MOT} Narrow-line MOT and dipole trap loading. The narrow-line cooling consists of five consecutive stages, labeled MOT I through V: a 50-ms capture MOT, during which the repumping of atoms from the $^3P_2$ to the $^1S_0$ state takes place, a compression MOT of 100\,ms, two cooling MOT stages of 100\,ms and 250\,ms, and a 100-ms hold time at final parameters. The end of the MOT phase is chosen as the origin of the time axis. (a) Intensity of the MOT beams in units of the saturation intensity $I_{\rm sat}=3\,\mu$W/cm$^2$. The intensity of the horizontal, upward, and downward MOT beams can be set independently. The downward MOT beam has little effect and is turned off before the other ones (see inset). (b) Frequency of the MOT beams, given as detuning from the Zeeman-shifted $\sigma^+$ transition. The light is frequency-modulated in the first two MOT stages. For clarity, only every tenth line of the resulting frequency comb is shown here. Atom number (c) and temperature (d) drop significantly during the end of MOT stage IV. The drop in atom number happens as the restoring force of the MOT becomes too weak to support atoms against gravity, and all atoms outside the dipole trap are lost. The last 100\,ms of hold time do not reduce the temperature further, but this stage is crucial to increase the density of atoms in the dipole trap.}
\end{figure*}

To load the red MOT, metastable state atoms are transferred into the $^1S_0$ ground state by optical pumping via the $5s5d\,{^3D_2}$ state. The temperature of the sample recovered from the metastable state reservoir is on the order of the Doppler temperature of the blue MOT, $T_{D,{\rm blue}}=720\,\mu$K, which is three orders of magnitude higher than the final temperature reached with the red MOT. We initially use red MOT parameters that allow to capture such a high temperature sample and then ramp these parameters to conditions in which a high phase space density is reached. The intensity and frequency of the MOT light during the capture phase and the subsequent ramps are given in Figs.~\ref{fig:red_MOT}(a) and (b). During the capture phase, the red MOT laser beams are frequency-modulated to increase the capture velocity of the MOT. In addition, the quadrupole magnetic field gradient is suddenly lowered to 1.15(5)\,G/cm in the vertical direction, which increases the capture volume, and then maintained at this value for the remainder of the experimental sequence. The detuning of the MOT light is reduced to compress the atomic cloud, and the intensity is reduced to lower the temperature. This provides optimal conditions for loading the reservoir dipole trap.

\textbf{Reservoir dipole trap ---} The reservoir dipole trap is a large-volume dipole trap consisting of a horizontally propagating beam. This beam is derived from a 5-W fiber laser operating at a wavelength of 1065\,nm with a linewidth of 0.5\,nm (IPG ytterbium fiber laser, model YLD-5-1064-LP). The beam contains up to 2\,W of power and is linearly polarized in the vertical direction. Cylindrical lenses are used to create a horizontally elongated beam profile. The beam has a vertical (horizontal) waist of $17.3(2)\,\mu$m ($298(3)\,\mu$m), yielding an aspect ratio of about 1:17. The trap frequencies are 600(6)\,Hz (35.3(4)\,Hz, 5.78(6)\,Hz) in the vertical (transverse horizontal, axial horizontal) direction at a power of 1760(35)\,mW as used in the experiment. Including gravitational sagging, the potential depth in the vertical direction is $\sim k_B\times9\,\mu$K. The errors given here are dominated by the uncertainty in the measurement of the laser beam powers, which we assume to be 2\%. The uncertainty in the trap oscillation frequencies is much smaller, and other effects, such as transmission losses of the glass cell and gravitational sagging in the dipole trap, are accounted for.

We increase the axial frequency slightly by a second, near-vertical beam with $297(3)\,\mu$m waist and circular polarization, using a second fiber laser of the same model as mentioned above. At the power of 740(15)\,mW used in the experiment, it has a trap depth of only $k_B\times0.29(1)\,\mu$K and increases the trap frequency in the $x$-direction to 8.14(8)\,Hz. This increased axial trap frequency slightly increases the density of the sample and reduces the timescale of BEC formation. It has no further effects, and otherwise identical results are obtained without this additional beam.

\textbf{Loading of the reservoir dipole trap ---} When the intensity of the MOT beams is low, gravity plays an important role in the MOT dynamics. The atoms are pulled towards the lower part of an ellipsoid of equal magnetic field magnitude located around the quadrupole magnetic field center. On this ellipsoid, the atoms are in resonance with the cooling light and levitated by it, giving the MOT a pancake shape, roughly matched by the shape of the reservoir dipole trap. The dipole trap is located $600(100)\,\mu$m below the center of quadrupole field, where the B-field has a magnitude of 75(10)\,mG and is oriented nearly vertically. To overlap the MOT with the dipole trap, we carefully adjust the MOT laser detuning. Figures~\ref{fig:red_MOT}(c) and (d) show the temperature and atom number evolution of the atomic cloud, while it is loaded into the reservoir dipole trap. At the endpoint of the ramp, the intensities of the horizontal (upward, downward propagating) MOT beams are $0.6\,I_{\rm sat}$ ($0.5\,I_{\rm sat}$, $0\,I_{\rm sat}$). The detuning from the $\sigma^+$-cooling transition is about $-20\,$kHz. With the given non-zero B-field at the position of the atoms, this corresponds to a detuning of about 150\,kHz from the unperturbed $\pi$-transition.

Laser cooling of the atoms in the dipole trap is influenced by Zeeman and light shifts. The vertical gradient of 1.15\,G/cm translates to a negligible shift of $600\,\mu$G across the sample, assuming an estimated vertical size of $5\,\mu$m. The horizontal gradient is 0.63\,G/cm, and the horizontal radius of the cloud is rather large: about $200\,\mu$m, giving a shift of 12\,mG between the center and the outsides, which corresponds to a significant frequency shift of 25\,kHz. The dipole trap induces a non-negligible differential AC Stark shift on the $^1S_0$ and $^3P_1$ states. We can reduce this shift by choosing the optimized polarizations for the horizontal and vertical dipole trap beams mentioned above, such that the magnitude of the shift is about 10\,kHz \cite{Boyd2007Thesis}. Assuming that the atoms explore one tenth of the trap depth, we obtain a shift across the sample of about 1\,kHz, which is negligible.

\begin{table*}[ht]
\caption{\label{tab:Tab1}Parameters of the dipole trap beams used for the experiments described in the Letter. Gravitational sagging is taken into account in the calculation of the horizontal dipole trap depth.}
\begin{ruledtabular}
\begin{tabular}{ccccccccc}
beam & waist $x$ & waist $y$ & waist $z$ & $P$ & $U/k_B$ & $f_x$ & $f_y$ & $f_z$\\
&  ($\mu$m) & ($\mu$m) & ($\mu$m) & (mW) & ($\mu$K) & (Hz) & (Hz) & (Hz) \\
\hline
horizontal & & 298(3) & 17.3(2) &  1760(35) & 9.2(2) & 5.78(6) & 35.3(4) & 600(6) \\
vertical & 297(3) & 297(3) & & 740(15) & 0.29(1) & 5.73(6) & 5.73(6) & $\sim 0$\\
dimple  & 22.4(2) & 22.4(2) & & 38.3(8) & 2.6(1) & 228(2) & 228(2) & $\sim 0$\\
\end{tabular}
\end{ruledtabular}
\end{table*}

About $9 \times 10^6$ atoms are captured in the reservoir dipole trap at a temperature of 830\,nK. The achievable temperature is density-dependent. A reduction of the atom number by a factor two leads to a temperature reduction of $\sim 100\,$nK. Over the course of one hour, the variation of atom number between experimental cycles is about 1\%, and the temperature variation is below 10\,nK. After loading of the dipole trap, the intensity of the upward cooling beam is reduced by a factor of 10, and the horizontal beams are turned off. Such a sample is the starting point for the subsequent laser cooling into quantum degeneracy.

\subsection{Transparency beam}

Immediately after completion of the dipole trap loading, the transparency beam is turned on. This beam is derived from a free-running master diode laser with a frequency stability of order 100\,MHz/day. The frequency is blue detuned by 15\,GHz from the $^3P_1 - {^3S_1}$ transition, and constantly monitored by a wavemeter. The beam is circularly polarized and focused onto the center of the atomic cloud with a waist of $26.2(3)\,\mu$m, again calculated from trap frequency measurements. The power used is 7.5(2)\,mW, translating to a peak intensity of $7 \times 10^5$\,mW/cm$^2 = 7$\,MW/m$^2$. The beam propagates downwards, at an angle of $15^{\circ}$ to vertical due to geometrical restrictions.

The transparency beam has a strong influence on the $^3P_1$ state. The differential AC Stark shift of the $^1S_0 - {^3P_1}$ cooling transition is on the order of $+10\,$MHz, such that the cooling light is red-detuned by $>1000\,\Gamma_{\rm red}$ for atoms at the center of the transparency beam. In this way, atoms illuminated by the transparency beam are transparent to laser cooling photons. Note that this scheme is drastically different to a scenario in which the cooling beam would contain a small dark spot imaged onto the dimple region: in this case, atoms in the dimple region could still absorb cooling light scattered by atoms in the reservoir. Our method differs also from the dark spot MOT technique, since we do not change the internal state of the atom in the region of transparency. A careful analysis of the magnitude of the AC Stark shift of the $^3P_1$ state is given in Sec.~\ref{sec:ESD}.

The transparency beam also creates an attractive trapping potential for the $^1S_0$ state with a depth of $k_B\times0.5\,\mu$K. Because of similar beam orientation and waists, the potential created by the transparency beam resembles the potential created by the dimple beam, but has only about 20\% of its depth.

\subsection{Cooling light}

The cooling light consists of an upward propagating, circularly polarized beam, red detuned by about 15\,kHz from the $\sigma^+$ $^1S_0 - ^3P_1$ transition. It has a power of 20\,nW and a peak intensity of $0.15\,\mu{\rm W/cm}^2$, corresponding to $0.05\,I_{\mathrm{sat}}$. The light has the same source as the light used for the narrow-line MOT. Any desired cooling rate of up to many 100\,nK/ms can be achieved by a suitable combination of detuning and intensity, where a larger detuning can be compensated by an increased intensity. We find a fixed relation between cooling rate and induced decay rate, which is independent on the combination of detuning and intensity. There is a lower temperature limit of about 450\,nK, below which cooling is accompanied by rapidly increasing loss rates. For the experiments described in the Letter, we use only one upward propagating cooling beam, and we find that the addition of further cooling beams from other directions does not improve the performance.

\subsection{Dimple}

The local increase in density is facilitated by the dimple beam. This beam is aligned almost vertically, with an angle of $22^{\circ}$ ($37^{\circ}$) towards vertical (the transparency beam) due to geometrical restrictions. It is derived from the same laser source as the vertical dipole trap, has circular polarization and a waist of $22.4(2)\,\mu$m. The dimple beam is centered in the plane of the reservoir trap with the transparency beam to within $5\,\mu$m. At a power of 38.3(8)\,mW, the dimple provides horizontal trap frequencies of 228(2)\,Hz and has negligible trap frequencies in the vertical direction. In presence of the dimple, we refer only to the region outside of the dimple as the reservoir.

The dimple trap is set to a small depth of $k_B \times 0.15\,\mu$K at the beginning of the experimental sequence and ramped in 10\,ms to a depth of $k_B \times 2.6\,\mu$K after the transparency beam is switched on. The ramp speed is not adiabatic with respect to the horizontal trap frequencies of the reservoir trap, and atoms from the reservoir continue to accumulate in the dimple for 10\,ms after the ramp. At this point, the temperature of the gas in the dimple is twice the temperature of the reservoir gas. Thermal contact with the laser-cooled reservoir lowers the temperature on a timescale of $\sim100\,$ms. In thermal equilibrium, the dimple leads to a peak density increase by a factor of 30 compared to the reservoir. This density increase is the origin of the gain in phase-space density.

The BEC phase transition is observed 50\,ms after ramping the dimple to high power. The BEC grows to slightly more than $10^5$\,atoms after another 100\,ms. The local harmonic potential confining the BEC is dominated horizontally by the dimple trap and vertically by the horizontal dipole trap, leading to trap frequencies of 228(2)\,Hz in the horizontal plane and 600(6)\,Hz in the vertical direction. The BEC is pancake-shaped with a horizontal (vertical) Thomas-Fermi radius of $5\,\mu$m ($2\,\mu$m).

\begin{figure}
\includegraphics[width=86mm]{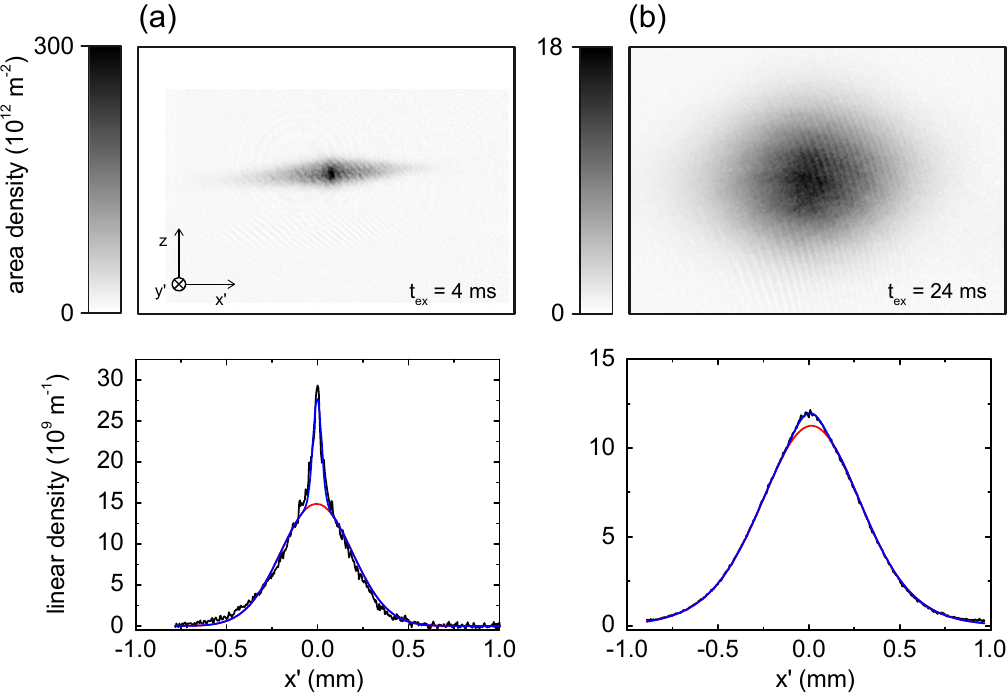}
\caption{\label{fig:ExampleExpansionPictures}Typical absorption images used to determine temperatures and atom numbers. (a) Absorption image after a short expansion time $t_{\rm ex}=4\,$ms. A two-dimensional double-Gaussian fit is used to extract atom number and temperature individually for dimple and reservoir. (b) Absorption image after a long expansion time $t_{\rm ex}=24\,$ms. A tri-modal two-dimensional fit, consisting of two Gaussians and an integrated Thomas-Fermi distribution, are used to extract the atom number of the thermal component and the BEC. The panels below the absorption images show the density distribution integrated along the $z$-direction (black) and the fits (blue). The red curves show the contributions of the reservoir (a) and of the thermal component (b).}
\end{figure}

\subsection{Absorption imaging and data analysis}

We use time-of-flight absorption images to deduce all relevant information from our atomic samples. We are interested in the following quantities: atom number in the dimple and in the reservoir, temperature of the dimple and of the reservoir, and number of atoms in the condensate.

The temperature and atom number of the dimple and reservoir regions are measured in absorption images taken after 4\,ms of free expansion; see Fig.~\ref{fig:ExampleExpansionPictures}(a). After this short time, the clouds from the two regions can still be clearly distinguished in the horizontal direction ($x^\prime$-direction), but have expanded well beyond their initial size in the vertical direction ($z$-direction) to allow thermometry. Since the atomic clouds are very dense and nearly opaque to resonant imaging light, we image at a detuning of $48\,\mathrm{MHz} \approx 1.5\,\Gamma$, which reduces the absorption cross section by a factor of about 12. We employ a 2D double-Gaussian fit to the data. The temperature is derived from the vertical expansion only.

The overall atom number and the BEC atom number are determined from absorption images taken after a free expansion time of 24\,ms; see Fig.~\ref{fig:ExampleExpansionPictures}(b). We fit the data with a 2D distribution consisting of two Gaussians for the thermal atoms and an integrated Thomas-Fermi distributions for the BEC. Two Gaussians are used to take into account the two different sources of thermal atoms, the reservoir and the dimple. If only one Gaussian is employed to describe all thermal atoms, the BEC atom number deduced from the Thomas-Fermi part of the fit increases by about 50\%.

All measurements are performed three times, and the average values and statistical error bars of atom number and temperature are given in Figs.~3 and 4 of the Letter.

\section{Characterization of laser-cooled BECs}
\label{sec:Characterization}

The BEC formation depends on various parameters. In this section, we will explore the influence of these parameters on the system and show that our method of laser cooling to quantum degeneracy works in a very broad range of parameters.

\subsection{BEC atom number}

\textbf{Total atom number and temperature ---} The lowest achievable temperature depends on the total atom number in the system as a consequence of red MOT dynamics. We measure the BEC atom number $N_{\rm BEC}$ in dependence of the total atom number $N$, at these lowest achievable temperatures. We find that $N_{\rm BEC} \propto N$. This behavior can be explained by the model described in Sec.~\ref{sec:Model}.

\textbf{Trap depth ---} We independently vary the trap depth of horizontal beam and dimple around the values presented in the Letter. A variation in trap depth of the horizontal beam by a factor of 1.7 changes the BEC atom number by at most 15\%, and a variation of dimple power by a factor of 2.5 changes the atom number by at most 30\%. A broad, global maximum of the BEC atom number exists, which is where we perform our experiments.

An important constraint on the trap depth is the desire to avoid evaporation. This constraint is fulfilled by the conditions chosen; see Sec.~\ref{sec:Evaporation} for details.

\textbf{Dimple ---} We find that the BEC atom number depends only weakly on the dimple size. We vary the waist of the dimple beam between 20 and $50\,\mu$m, but find no appreciable change in BEC atom number. The BEC formation time, however, is increased for larger waists. We also vary the ramp-up time of the dimple between 0 and 100\,ms, but find no influence on temperature or BEC atom number after a 250-ms equilibration time.

\subsection{Transparency beam}

The transparency beam is the key novelty of our work. A detailed study of the non-trivial shifting of the laser cooling transition can be found in Sec.~\ref{sec:ESD}.

\textbf{Frequency ---} The transparency beam is employed to locally shift the energy of the $^3P_1$ state by about $+10\,$MHz. This would in principle be possible with light blue-detuned to any transition originating from the $^3P_1$ state. We chose to work in the vicinity of the $^3P_1 - {^3S_1}$ transition at 688\,nm for two reasons: The availability of diode lasers, and the negligible influence on the $^1S_0$ state.

In our experiment, the transparency beam is blue-detuned by about 15\,GHz from the $^3P_1 - {^3S_1}$ transition. In a series of measurements, we set the detuning to different values and vary the intensity of the transparency beam while searching for the appearance of a BEC. We never observe BEC formation for frequencies around or red-detuned to the $^3P_1 - {^3S_1}$ transition. We observe BEC formation only in a frequency range between 6 and 30\,GHz blue detuning, where the upper bound of 30\,GHz is probably limited by the available laser power of 10\,mW.

\textbf{Intensity ---} We observe the formation of a BEC only above a certain critical intensity $I_c$ of the transparency beam, which depends on its frequency. The BEC atom number quickly grows for larger intensities, and saturates at about $2\,I_c$. The experiment is performed at around $10\,I_c$, corresponding to a peak intensity of $0.7\,$kW/cm$^2$.

\textbf{Spectral filtering ---} The $^3P_1 - {^3S_1}$ transition is 1.428\,nm (or 902\,GHz or 30\,cm$^{-1}$) away from the $^1S_0 - {^3P_1}$ intercombination transition, and off-resonant excitation of the intercombination transition is negligible. The light originating from the laser diode, however, contains incoherent fluorescence light, also known as residual amplified spontaneous emission (ASE). This light covers a broad spectrum with a width of about 20\,THz, which includes the $^1S_0 - {^3P_1}$ transition. A spectrometer is used to estimate the spectral power of the incoherent background $P_{\rm{ASE}}$ to be on the order of $10^{-12}P_0$ in a 1\,kHz wide band 1\,THz away from the laser line, where $P_0$ is the power in the laser line. The scattering of these incoherent photons limits the lifetime of a BEC to 800\,ms when illuminated by unfiltered transparency light. Spectral filtering of the light allows us to reduce the amount of resonant light by a factor of 500, leading to an increased lifetime of 10\,s. The lifetime measurement is performed with a pure BEC obtained by standard evaporation, and its lifetime is limited to 10\,s by inelastic collisions.

\textbf{Waist ---} The transparency beam needs to be well-aligned with the dimple beam, and it needs to cover the BEC entirely. We vary the waist of the transparency beam between 10 and $55\,\mu$m. For comparison, the dimple waist is $22\,\mu$m and the Thomas-Fermi radius of the BEC in the $xy$-plane is $5\,\mu$m. Given an appropriate adjustment of the intensity, we can create laser-cooled BECs within the entire range of transparency beam waists examined. The purification method described in Fig.~2(c), during which reservoir atoms are selectively removed, becomes increasingly inefficient for larger beam sizes, as also parts of the reservoir are shielded. A larger beam size makes the system insensitive to misalignments of the beams, which we verify experimentally.

\textbf{Local density increase ---} One might speculate that the transparency beam leads to an increase in atom density purely by elimination of photon emission and re-absorption cycles, which act as an effective repulsion. This is not the case. We do observe a small density increase induced by the transparency beam, but it can be explained entirely by its trapping potential for atoms in the $^1S_0$ state.

\subsection{Timescales}

\textbf{Collision rate and thermalization ---} The peak scattering rate of thermal atoms in the dimple region is $3100\,\rm{s}^{-1}$, found at the edge of the BEC (see Sec.\ref{sec:Model}). The scattering rate decreases with distance from the center of the dimple, down to $50\,\rm{s}^{-1}$ outside of the dimple region. We observe that thermalization happens on a timescale of about 100\,ms, which corresponds to about 5 collisions; compare Fig.~3(b). This number matches well with the expectation of about 3 collisions required for thermalization.

\textbf{BEC formation time ---} We study the time required for the BEC to form after the dimple has been ramped up. For a BEC to be created, the atom number in the dimple has to be high enough and the temperature below a critical temperature for the given atom number. We find that the timescale required to accumulate atoms in the dimple is much shorter than the timescale required to thermalize the dimple gas with the reservoir; see Fig.~3. Thus, the BEC formation time is limited by the thermalization time scale. This time scale is linked to the elastic collision rate, which depends on the density of the sample. To test this relation, we vary the density by two means, changing the reservoir atom number and changing the reservoir trap frequency. We observe an increase in formation time as we decrease the atom number, down to a minimum atom number of about $1\times10^6\,$atoms, below which no BEC forms in our trap geometry. Using the trap parameters of the experiment described in the Letter, the fastest formation time is about 50\,ms for the maximum achievable atom number of $10\times10^6\,$. An increase of the axial frequency from 8\,Hz to 40\,Hz reduces the formation time to about 20\,ms. The timescale of BEC formation (following abrupt changes of the thermal distribution) was studied in previous experiments \cite{Kohl2002gob,Ritter2007otf} and found to depend on the scattering rate, in agreement with our measurements.

\textbf{Reservoir lifetime ---} A sample of $9\times10^6$ atoms at 800\,nK confined in the horizontal dipole trap has a lifetime of $\sim 30\,$s. When adding the dimple beam to obtain settings identical to the ones used in the experiment described in the Letter, the lifetime is reduced to 3.4(1)\,s. We believe that this lifetime is mainly limited by 3-body collisions in the dimple region, as the reservoir constantly replenishes the dimple population. The cooling light induces an additional decay, which is strongly dependent on intensity and detuning. For the values used in the experiment, the lifetime of the reservoir is reduced to about 2.5\,s.

\textbf{BEC lifetime ---} We observe that the lifetime of our BEC is linked to the lifetime of the reservoir gas. The reason is that any atom loss from the BEC is quickly and continuously replaced by atoms from the reservoir. A BEC exists as long as the total atom number in the system is above $\sim 10^6$ atoms. In order to measure the bare lifetime of the BEC without constant replenishment, we create a BEC, remove the atoms in the reservoir, and turn off the cooling light and the transparency light. The lifetime of this BEC is 1.0(1)\,s, probably limited by heating and loss resulting from inelastic 3-body collisions. The same result is obtained for an identical BEC created by conventional evaporation and re-compressed
into the dimple. By contrast, the lifetime of a BEC in presence of the reservoir, the transparency light, and the cooling light corresponds to the reservoir lifetime of 2.5\,s.

\textbf{Performance of the transparency beam ---} The transparency beam is absolutely necessary to protect the BEC from destruction by the cooling light. With the transparency beam turned off and the BEC subjected to cooling light, we measure a decay rate of about 1\,ms, and the BEC is completely destroyed in less than 3\,ms.

To demonstrate the ability of the transparency light to protect the BEC from resonant photons, we measure the lifetime of a BEC after removal of the reservoir and in presence or absence of both, the transparency and the laser cooling light. The lifetime is the same in both cases, 1\,s. We conclude that the BEC lifetime is not limited by scattering of laser cooling photons if the transparency light is present. We give an upper bound of $1\,$s$^{-1}$ for the scattering rate of laser cooling photons, which is consistent with measurements presented in Sec.~\ref{sec:ESD}.

The transparency beam itself has no negative effect on the BEC: lifetime measurements let us deduce an absorption rate of photons from this light field to be well below 0.1\,s$^{-1}$.

\textbf{Vacuum lifetime ---} Limitations to the lifetime of ultracold atoms in a dipole trap (beyond 3-body losses and residual evaporation) are off-resonant scattering of dipole trap photons, background gas collisions, and resonant stray light. We measure the lifetime of atoms at low density in a deep dipole trap to be 120\,s, much longer than any other timescale. Note that all of the lifetimes evaluated here are between one and three orders of magnitude larger than the BEC formation time.

\section{Absence of evaporation}
\label{sec:Evaporation}

\begin{figure*}
\includegraphics[width=179mm]{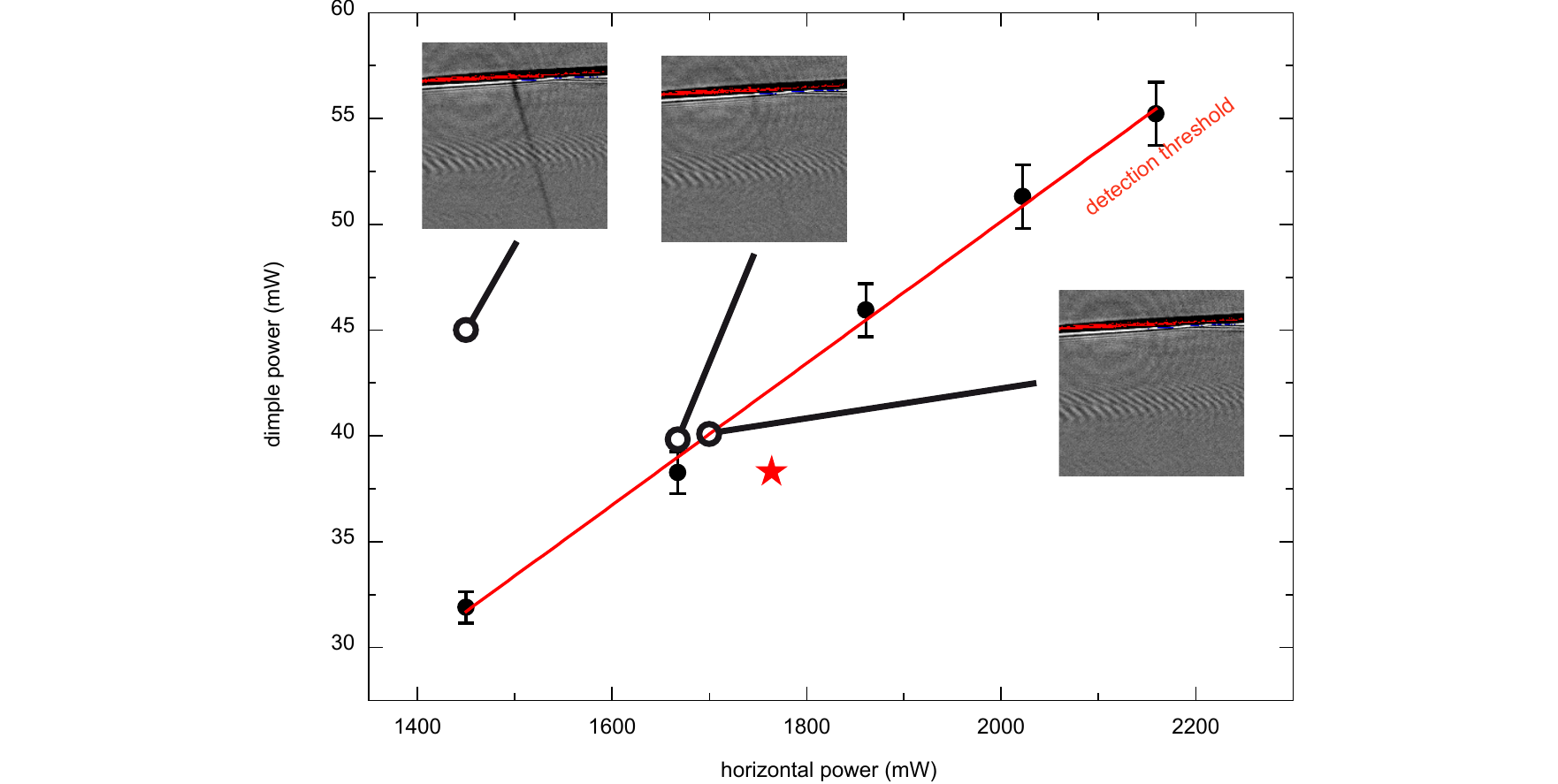}
\caption{\label{fig:evaporation} Detection of residual evaporation. Hot atoms from the dimple region might leave the dipole trap along gravity. When doing so, they are funneled into the dimple beam and can be detected by absorption imaging (insets). The amount of evaporation depends on the power in the horizontal and dimple trap beams, illustrated by the three insets showing atom fluxes of 350, 120, and $<50$\,atoms/ms. Our detection limit is about 50\,atoms/ms, and conditions with atom fluxes that touch this limit are indicated by the black data points. The straight line is a linear fit to the data. Experiments are performed in the region below this line, indicated by the star. The images are averages of 10 experimental realizations, and the color scale is adjusted to maximize contrast of small signals.}
\end{figure*}

The novelty of our work lies in the fact that BEC is achieved purely by laser cooling and thermalization between atoms, and not by evaporation and associated loss to the outside environment. Standard evaporative cooling of $^{84}$Sr has been used in past experiments and can be very efficient: for a factor of 10 loss in atom number, more than three orders of magnitude in phase space density can be gained \cite{Stellmer2009bec}. Thus, already little evaporative atom loss can increase the phase space density considerably. It would require an infinitely deep trap to rigorously exclude evaporation. Here, we quantify the flux of evaporated atoms and ensure that evaporation is reduced to a negligible level.

We observe an overall atom loss that leads to a lifetime of the sample of about 3\,s. This loss is not a measure of evaporation, as it includes various other loss processes as well: 3-body recombination, loss by scattering of cooling photons, light-assisted collisions, and off-resonant optical pumping into the $^3P_{0,2}$ states by the transparency light. We therefore need a specific probe for evaporated atoms.

It is also important to understand that the temperature of atoms in the reservoir is set entirely by the cooling light. This is true even if efficient evaporation would be taking place in the reservoir: in this case, the cooling light would heat the sample. In the following, we will be concerned only about evaporation in the dimple region, where the sample is transparent to cooling light.

The atoms are supported against gravity by the horizontal trap, and further confined by the dimple beam, which does not support the atoms against gravity. Atoms from the horizontal beam will preferentially evaporate downwards, aided by gravity. If they do so within the dimple region, they are attracted by and guided in the dimple beam. This beam provides superb confinement of evaporated atoms and enables us to detect even minute atom numbers in absorption imaging. The dimple beam has a potential depth of $k_B\times 2.6\,\mu$K and therefore selectively guides only low-energy atoms from evaporation, but not atoms originating from inelastic collisions.

The amount of atoms leaking from the dimple region into the dimple beam will depend on the sample temperature, the power of the horizontal beam, and the power of the dimple beam. We set the temperature to the one used in the experiment, vary the power of the horizontal and the dimple beam independently, and measure the atom leakage. The result of such a measurement is shown in Fig.~\ref{fig:evaporation}. The data points denote combinations of parameters where the atom flux is about 50\,atoms/ms, which is our detection threshold. The straight line is a linear fit to the data. Note that for a leakage of 50\,atoms/ms and a BEC formation time below 60\,ms, less than 3\,000 out of $9\times10^6\,$atoms evaporate in the dimple region. For the experiments described in the Letter we verify that no discernable evaporation takes place at any time during the optical cooling.

We explore the entire parameter region shown in Fig.~\ref{fig:evaporation} and find no dramatic difference in BEC formation time or final BEC atom number, again indicating that any residual evaporation has no measurable contribution to the BEC formation.

\section{Shifting atomic and molecular transitions}
\label{sec:ESD}

\subsection{Introduction}

We intend to cool atoms in the dipole trap with light resonant to the $^1S_0 - {^3P_1}$ transition, but only in the region outside the dimple. To do so, we render the atoms in the dimple region transparent for the cooling light by locally shifting the $^3P_1$ state out of resonance. Cooling of atoms in the reservoir is performed on the red side of the $^1S_0 - {^3P_1}$ transition, so we need to locally shift the $^3P_1$ state towards higher energies by at least a few 10 linewidths. This is done with the transparency beam, which spatially overlaps with the dimple region.

\begin{figure}
\includegraphics[width=86mm]{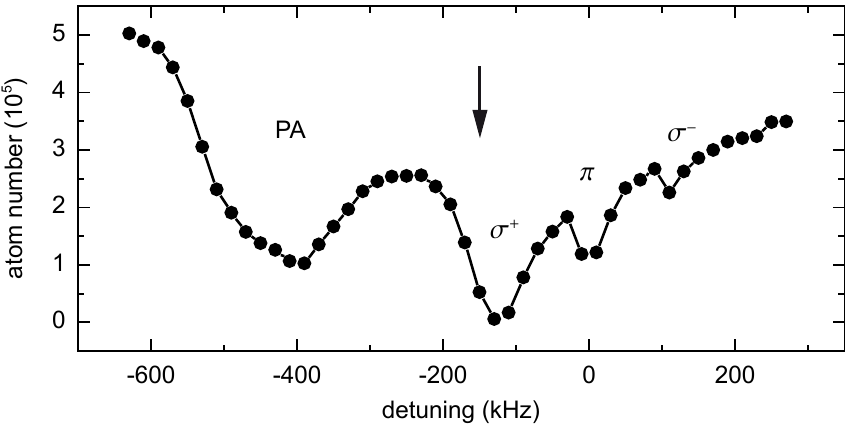}
\caption{\label{fig:example_spectrum} Loss spectrum of atoms confined in a dipole trap and illuminated by $\sigma^+$-polarized cooling light. A small guiding field of 90\,mG splits the $^3P_1$ state into three clearly visible $m_J$ states. The broad feature around $-400$\,kHz is due to photoassociation (PA). The visibility of $\pi$ and $\sigma^-$ transitions indicates that the cooling light is not perfectly circularly polarized. In the experiment described in the Letter, cooling is performed on the red side of the $\sigma^+$ transition, indicated by the arrow. The line is a guide to the eye. Note that the detuning here is calculated from the resonance position at zero B-field, which is determined to better than 10\,kHz.}
\end{figure}

To characterize the system, we work in similar conditions as used in the experiment described in the Letter. Atoms are tightly confined in a crossed-beam dipole trap, consisting of the horizontal trap and the dimple, and we evaporate into a pure BEC to reach both high densities (to enhance possible photoassociative loss) and low trap depths (such that atoms can leave the trap with very few photon recoils). We then illuminate the atoms with circularly polarized cooling light, using only an upward propagating beam. The intensity of the cooling light is $1.2\,I_{\rm sat}$ and the duration of illumination is 10\,ms. A magnetic field of 90\,mG magnitude and vertical orientation is present. A typical loss spectrum is shown in Fig.~\ref{fig:example_spectrum}; note the broad photoassociation (PA) feature, which corresponds to the $\nu=-1$ state of the $1(0^+_u)$ potential \cite{Stellmer2012cou}.

\subsection{Shifting of resonances}

\begin{figure*}
\includegraphics[width=179mm]{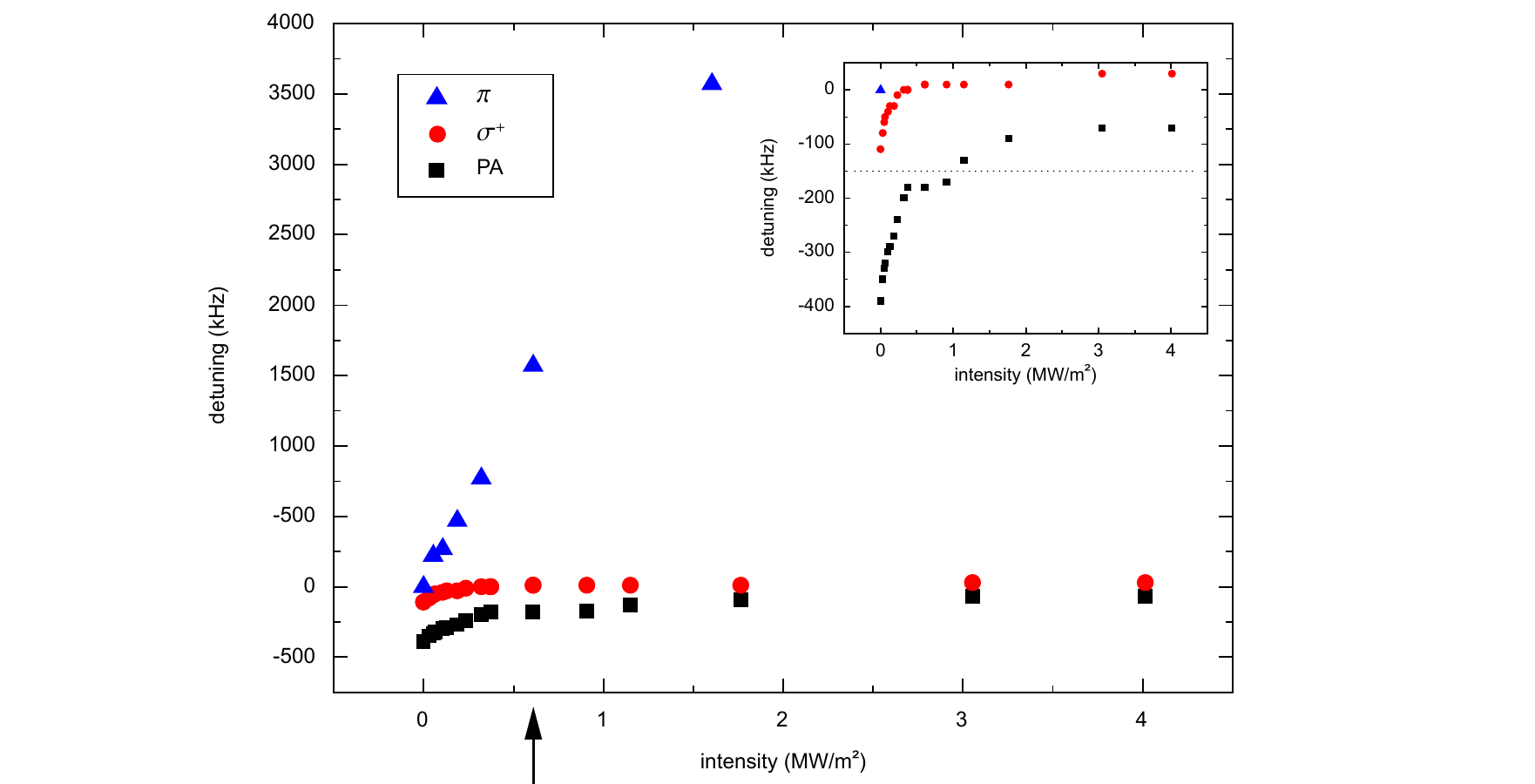}
\caption{\label{fig:linear_ESD} Resonance positions in dependence of transparency light intensity. Spectra as the one shown in Fig.~\ref{fig:example_spectrum} are taken for various intensities of the transparency beam, and the resonance positions recorded. The $\sigma^-$ transition is too weak to be tracked. Inset: a magnification of the region around the cooling frequency, denoted by the dotted line. An intensity of 0.6\,MW/m$^2$ (arrow) is used to take the data presented in Fig.~\ref{fig:polarization_ESD}.}
\end{figure*}

In a next step, we illuminate the atoms with transparency light tuned 15\,GHz blue to the $^3P_1 - {^3S_1}$ transition. The light is linearly polarized, propagating at an angle of $15^{\circ}$ to vertical. The waist is set to $35\,\mu$m to ensure a uniform light intensity across the sample. We take loss spectra as the one shown in Fig.~\ref{fig:example_spectrum} for various intensities and observe that the four loss features experience different Stark shifts and move differently with intensity; see Fig.~\ref{fig:linear_ESD}. The $\pi$-transition moves rapidly towards blue detuning, whereas the $\sigma^+$ and PA lines approach asymptotic values less than 100\,kHz from resonance.

For our cooling strategy, we intend to work at a detuning of about $-150$\,kHz from the resonance frequency at zero B-field. It seems possible to move the $\sigma^+$ transition a few 100\,kHz away from this frequency. Unfortunately, the broad PA transition moves into exactly this region, potentially causing undesired losses given the high density in the dimple region. The transitions broaden considerably with increased transparency beam intensity.

\subsection{Polarization dependence}

\begin{figure*}
\includegraphics[width=170mm]{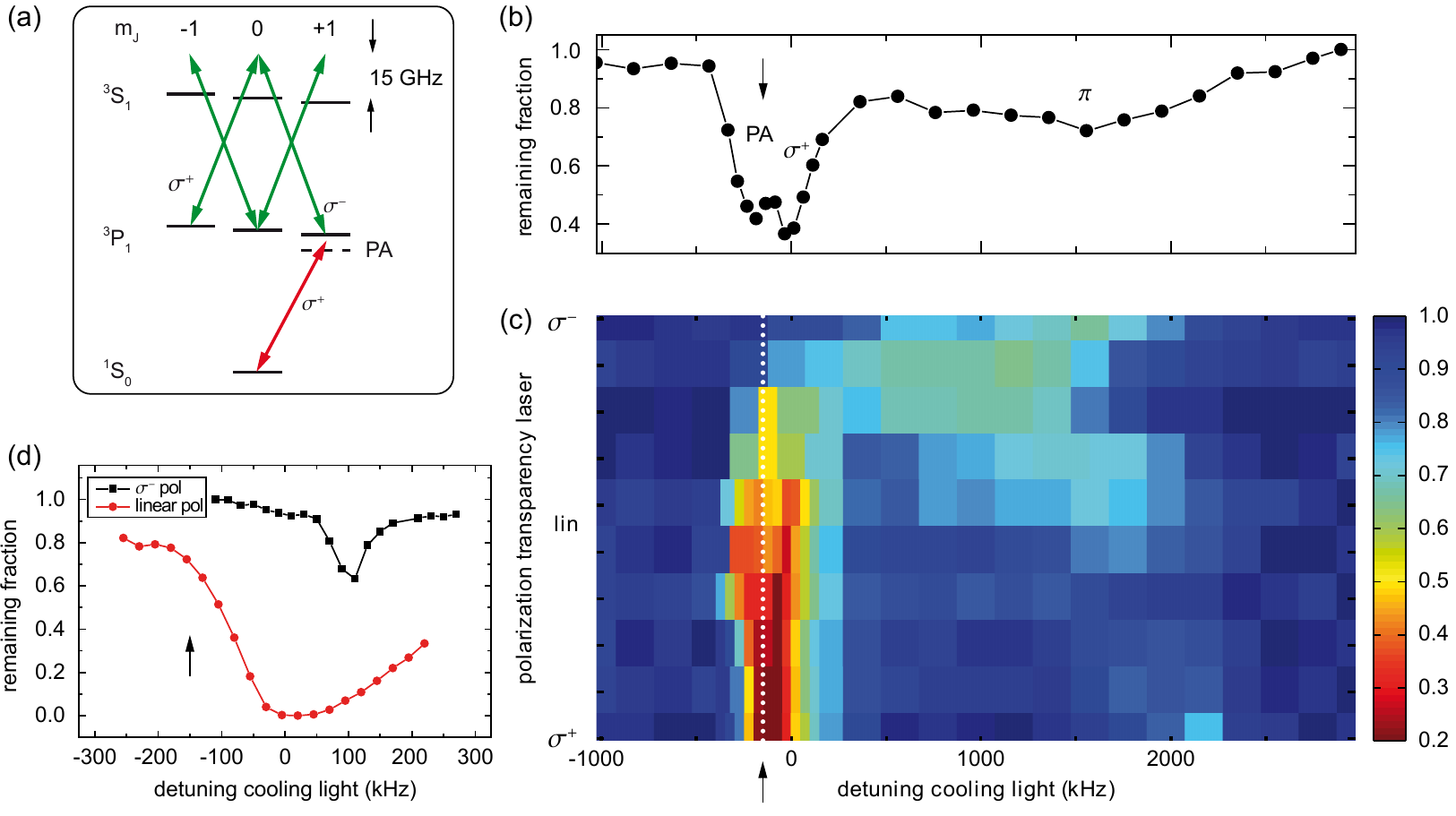}
\caption{\label{fig:polarization_ESD} Resonance positions in dependence of the polarization of the transparency light. (a) The three $m_J$ states of the $^3P_1$ manifold couple differently to light on the $^3P_1 - {^3S_1}$ transition (green arrows). The $m_J=+1$ state, for example, does not couple at all to $\sigma^+$-polarized light due to the absence of any $m_J=+2$ state in the $^3S_1$ manifold. The cooling light (red arrow) is always $\sigma^+$-polarized, with small admixtures of other components due to experimental imperfections. (b) A loss spectrum of atoms illuminated by the transparency light and subjected to a pulse of cooling light. The frequency of the cooling light is scanned, and the number of remaining atoms is recorded. The three loss features correspond to the PA-, $\sigma^+$-, and $\pi$-transitions. This spectrum is taken with a near-linear polarization of the transparency light and corresponds to the data set indicated by an arrow in Fig.~\ref{fig:linear_ESD}. (c) We take spectra at various polarizations, covering the complete range between $\sigma^+$-circular (bottom) to linear and $\sigma^-$-circular  polarization (top). The fraction of remaining atoms is encoded in the color, with red color denoting high atom loss. The dotted white line indicates the frequency of the cooling light. (d) Loss spectra for two extreme values of the polarization, taken with the intensity of the transparency light used in the experiment described in the Letter, and applying a 25-times stronger cooling pulse compared to the measurements shown in (b) and (c). The spectra taken with intermediate polarizations (not shown) smoothly interconnect between these two curves. The cooling frequency is again indicated by an arrow.}
\end{figure*}

So far, we have not considered the polarization of the transparency beam, which was linear in the previous measurement. This beam is almost co-aligned with the magnetic quantization axis, such that its field can be decomposed into equal amounts of $\sigma^+$ and $\sigma^-$-light (but no $\pi$-light) in the reference frame of the atoms. The three $m_J$ states of the $^3P_1$ manifold, as well as the PA line tied to the $m_J=+1$ state, couple differently to the $^3S_1$ states; see Fig.~\ref{fig:polarization_ESD}(a). This helps to understand the different shifts of the $\pi$- and $\sigma^+$-transitions in Fig.~\ref{fig:linear_ESD}, and we expect the shifting of the individual states to strongly depend on the polarization of the transparency light.

In a next step, we fix the intensity of the transparency light intensity to 0.6\,MW/m$^2$ and take scans at various polarizations. The results are shown in Figs.~\ref{fig:polarization_ESD}(b) and (c). The loss features change dramatically in position, width, and amplitude depending on the polarization of the transparency beam. The $\sigma^+$-transition and its associated PA line are maximally shifted by a $\sigma^-$-polarized transparency beam. This can easily be understood from the fact that $\sigma^+$-polarized light cannot couple the $m_J=+1$ state to the $^3S_1$ manifold. We find that already small variations in the polarization decrease the performance considerably. Note that in the experiment described in the Letter, we use an intensity of the transparency light that is larger by a factor of about 10.

We will now focus on the region of interest around the cooling frequency and highlight the importance of the polarization of the transparency light under conditions close to the ones used in the experiment described in the Letter. We increase the intensity of the transparency beam to $7\,$MW/m$^2$ and make the cooling pulse 25-times stronger (5-times longer and 5-times more intense) to amplify even small losses. Two spectra for $\sigma^-$ and near-linear polarization are shown in Fig.~\ref{fig:polarization_ESD}(d). Not only does the loss feature move to positive detuning with increasing circular polarization, also its width and amplitude decrease dramatically. The small remaining loss feature is about $250\,{\rm kHz} = 35\,\Gamma$ away from the desired cooling frequency. We speculate that this is the $m_J=-1$ state, which is unaffected by transparency light of $\sigma^-$-polarization and driven by small $\sigma^-$-polarization components in the cooling light.

\subsection{Conclusion}

We have shown that the magnetic substates of the $^3P_1$ state, as well as the highest molecular level of the $1(0^+_u)$ potential, can be shifted by a light field slightly detuned from the $^3P_1 - {^3S_1}$ transition. In particular, we find that the three Zeeman lines and the first PA line move very differently in dependence of the polarization of the transparency laser. We find that all lines can be moved at least 250\,kHz away from the desired cooling transition, reducing the photon scattering rate by at least three orders of magnitude. The strongest lines are even moved by a few MHz. This transparency beam allows us to cool atoms in the reservoir, while keeping the photon scattering rate of atoms in the dimple below the rate of 3-body collisions, which eventually determine the lifetime of the BEC. Further investigations are needed to gain a quantitative understanding of the AC Stark shifts in dependence on transparency laser frequency, intensity, and polarization.

\section{Density distribution models}
\label{sec:Model}

We now determine the density distributions of our samples by a numerical model and by a simple analytical model. Both models assume that the gas is in thermal equilibrium and neglect the influence of laser cooling light on the density distribution in the reservoir. The models are compared to the experiment and used to determine important quantities, such as the elastic collision rate.

\subsection{Self-consistent mean-field equations}

We use a self-consistent mean-field model to describe the density distribution of the BEC $n_0(\textbf{r})$ and of the thermal gas $n_{\rm th}(\textbf{r})$. The thermal gas is described as a gas of non-interacting particles residing in the potential $U_{\rm th}(\textbf{r})=U_{\rm ext}(\textbf{r})+2g(n_0(\textbf{r})+n_{\rm th}(\textbf{r}))$, which is the sum of the external trapping potential $U_{\rm ext}$ and the mean-field potential of all atoms. Here $g=4\pi\hbar^2 a/m$, where $a$ is the scattering length and $m$ the mass of the atoms. The potential $U_{\rm ext}$ is assumed to be zero at its deepest point. In the semiclassical approximation, the thermal density distribution in dependence of the potential is given by

\begin{equation*}
n_{\rm th}(U)=\lambda_{\rm th}^{-3} g_{3/2}(e^{-[U-\mu]/k_B T}),
\end{equation*}

\noindent where $\lambda_{\rm th}=h/\sqrt{2 \pi m k_B T}$ is the thermal de Broglie wavelength and $\mu$ the chemical potential. The function $g_{3/2}(z)$ is a polylogarithm function, $g_\alpha(z)=\sum_{n=1}^\infty z^n/n^\alpha$. The spatial density distribution is then

\begin{equation} \label{Eqn:Calc:ThermalDensity}
n_{\rm th}(\textbf{r})=n_{\rm th}(U_{\rm th}(\textbf{r})).
\end{equation}

\noindent The BEC distribution is obtained in the Thomas-Fermi (TF) approximation

\begin{equation} \label{Eqn:Calc:BECDensity}
n_0(\textbf{r})=\max[(\mu-U_0(\textbf{r}))/g,0],
\end{equation}

\noindent where the potential $U_0(\textbf{r})=U_{\rm ext}(\textbf{r})+2 g n_{\rm th}(\textbf{r})$ is the sum of external potential and the mean-field potential of the thermal atoms.

\subsection{Numerical model}

To determine the density distributions $n_0$ and $n_{\rm th}$, we self-consistently solve Eqns.~\ref{Eqn:Calc:ThermalDensity} and \ref{Eqn:Calc:BECDensity}. The solution is obtained by a numerical calculation for given values of the chemical potential $\mu$ and the temperature $T$. The calculation starts from initial density distributions that are zero everywhere, and obtains better and better approximations for the distributions by iterating Eqns.~\ref{Eqn:Calc:ThermalDensity} and \ref{Eqn:Calc:BECDensity}. The potential is described by a sum of the gravitational potential and the potentials created by the dipole trapping beams, which we assume to be Gaussian.

To speed up calculations, we first calculate tables containing $n_0(U)$ and $n_{\rm th}(U)$ for a range of relevant potential depths $U$. We then interpolate these tables to obtain $n_0(U)$ and $n_{\rm th}(U)$ for any $U$. The spatial density distributions $n_0(\textbf{r})$ and $n_{\rm th}(\textbf{r})$ are obtained in the local density approximation using realistic potentials $U(\textbf{r})$. To avoid numerical artifacts, the self-consistent solution of Eqns.~\ref{Eqn:Calc:ThermalDensity} and \ref{Eqn:Calc:BECDensity} is obtained in two stages. During the first stage, the mean-field of the thermal atoms is not included in the potential that the thermal atoms experience. After 20 rounds of iteratively solving Eqns.~\ref{Eqn:Calc:ThermalDensity} and \ref{Eqn:Calc:BECDensity}, we linearly ramp on this potential contribution in a second stage consisting of 25 rounds. In every round, the minimum $U_{\rm th}^{\rm min}$ of $U_{\rm th}$ is determined. Since the presence of a BEC is assumed, $n_{\rm th}(U_{\rm th}^{\rm min})$ equals $n_c$, which is used as a boundary condition in the determination of $n_{\rm th}(\textbf{r})$. To avoid oscillations in the calculation, we low-pass filter $U_{\rm th}^{\rm min}$ and use the filtered value in the calculation of $n_{\rm th}(\textbf{r})$.

In the experiment, we determine the atom number in the reservoir $N_{\rm res}$ and the atom number in the dimple $N_{\rm dimple}$ by performing a double-Gauss fit to the density distribution recorded after a short expansion time. To determine similar values from the numerical calculation, we define the reservoir density distribution $n_{\rm th, res}(\textbf{r})=n_{\rm th}(U_{\rm res}(\textbf{r}))$, where $U_{\rm res}(\textbf{r})$ is a potential without the contribution of the dimple, but the same potential values outside the dimple region. The atom numbers in BEC and reservoir are determined by integrating $n_0(\textbf{r})$ and $n_{\rm th, res}(\textbf{r})$, respectively. The thermal atom number in the dimple is obtained by integrating $n_{\rm th, dimple}(\textbf{r})=n_{\rm th}(\textbf{r})-n_{\rm th, res}(\textbf{r})$. The radii $R_i$ ($i\in\{x,y,z\}$) of the BEC are defined by the locations at which the BEC density drops to zero.

An important quantity which we want to extract from our model is the elastic collision rate of the thermal gas. It is given by $\Gamma_{\rm el,th}(n)=n v \sigma$, where $v=4\sqrt{k_B T/ \pi m}$ and $\sigma=8 \pi a^2$. The peak collision rate of the thermal gas $\Gamma_{\rm el, th, peak}$ is reached at the edge of the BEC, where the gas has the critical density. The timescale of thermalization between the gas in the dimple and the reservoir gas is limited by the elastic collision rate in the reservoir, outside the dimple region. This rate is highest just to the side of the dimple in the $xy$-plane, and determined from $n_{\rm th}$ at that location, which we denote as $n_{\rm th, edge}$. Assuming a Gaussian reservoir density profile in the $z$-direction and a homogeneous distribution in the $xy$-plane, the average elastic collision rate in the reservoir close to the dimple is $\Gamma_{\rm el, th, res}=n_{\rm th, edge} v \sigma /\sqrt{2}$.

\subsection{Analytical model}

Before we present the results of the numerical model, we discuss a simple analytical model. We neglect the mean-field of the thermal atoms and approximate the dimple potential by a harmonic potential with trap oscillation frequencies $f_i$. The BEC resides entirely in the dimple and has the usual TF inverted parabola shape. The chemical potential is determined from the BEC atom number $N_0$ by

\begin{equation*}
\mu=\frac{\hbar \omega}{2}\left( \frac{15 N_0 a}{a_{\rm ho}}\right)^{2/5} ,
\end{equation*}

\noindent where $a_{\rm ho}=\sqrt{\hbar/m \omega}$ and $\omega=2\pi (f_x f_y f_z)^{1/3}$. The radii of the BEC are given by the TF radii $R_{{\rm TF},i}=\sqrt{2\mu/m (2\pi f_i)^2}$ and the peak density is $n_0(0)=\mu/g$.

Because of different correlations in the BEC and the thermal gas, the BEC mean-field is twice as strong for thermal atoms than for atoms in the BEC. Thermal atoms in the center of the trap experience a ``Mexican hat'' potential, created by the external confinement combined with the BEC mean-field. In the center of the BEC, this mean-field reaches $2 g n_0(0)=2\mu$. For our typical parameters (see Tab.~\ref{tab:Calc:Tab2}), the chemical potential is on the order of half the temperature, $\mu \sim 0.5 k_B T$, which means that the thermal distribution is strongly modified by the BEC mean-field. The density of the thermal gas reaches the critical density $n_c=n_{\rm th}(\mu)=2.612/\lambda_{\rm th}^3$ at the edge of the BEC. The dimple potential has a finite size in the $xy$-plane and a finite depth $U_{\rm dimple}^{\rm max}$. At the edge of the dimple, the density is $n_{\rm th, edge}=n_{\rm th}(U_{\rm dimple}^{\rm max})$. The elastic collision rate in the reservoir is given by $\Gamma_{\rm el,th,res}=\Gamma_{\rm el,th}(n_{\rm th, edge})$ and the peak elastic collision rate of the thermal gas is again  $\Gamma_{\rm el, th, peak}=\Gamma_{\rm el,th}(n_c)$.

We can estimate the atom number in the reservoir $N_{\rm res}$ by describing the reservoir as a harmonic potential with trap oscillation frequencies $f_{{\rm res}, i}$. The central density of the reservoir is about  $n_{\rm th, edge}$, which is much lower than the critical density. Therefore the gas in the reservoir is well described by a classical thermal distribution. With this approximation, we obtain $N_{\rm res}=n_{\rm th, edge} (2\pi k_B T/m)^{3/2}/\omega_{\rm res}^3$, where $\omega_{\rm res}=2\pi(f_{{\rm res}, x} f_{{\rm res}, y} f_{{\rm res}, z})^{1/3}$.

\begin{figure*}
\includegraphics[width=179mm]{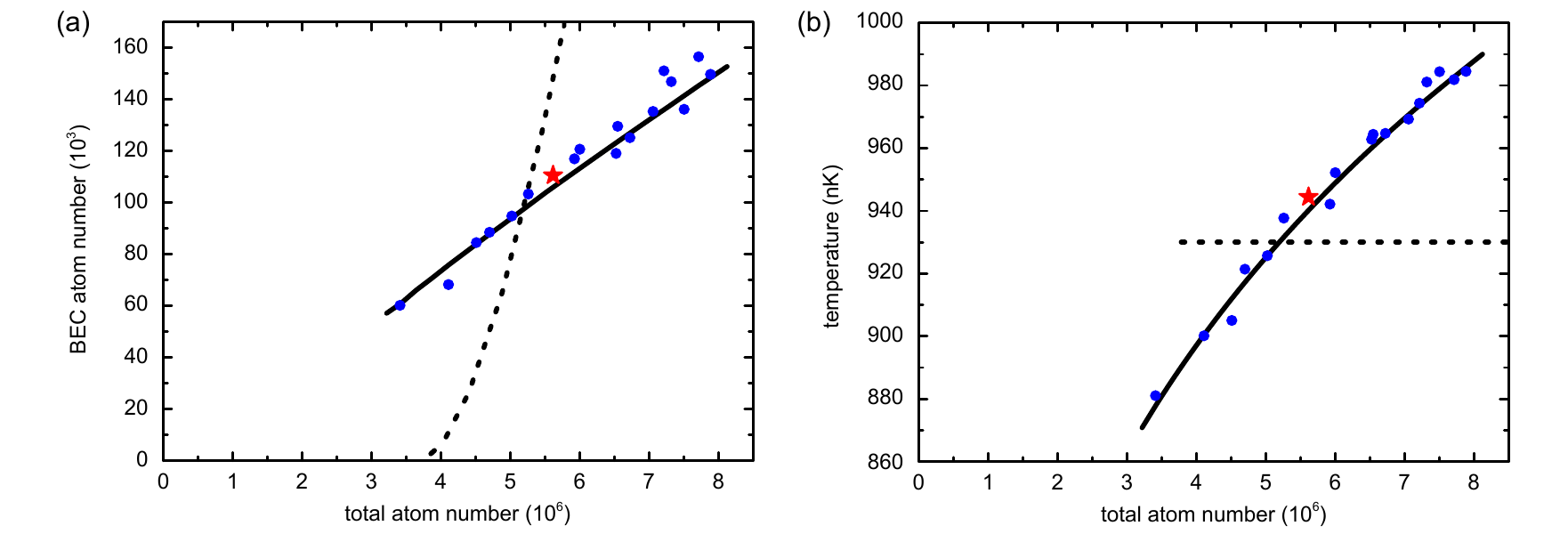}
\caption{\label{fig:Calc:Fig1} Comparison of experiment (blue data points) and numerical model (solid lines). Shown are BEC atom number (a) and temperature (b) in dependence of the total atom number. Experiment and calculation agree well. A peculiar feature is that the BEC atom number is nearly proportional to the total atom number. An important contribution to this behavior is the increase in temperature with the atom number observed in the experiment. For comparison, a calculation for constant temperature is shown as dashed line. Here, the BEC atom number increases much more steeply with total atom number. Results of the models matched to the star-shaped data point are given in Tab.~\ref{tab:Calc:Tab2} and Fig.~\ref{fig:Calc:Fig2}.}
\end{figure*}

\begin{figure*}
\includegraphics[width=179mm]{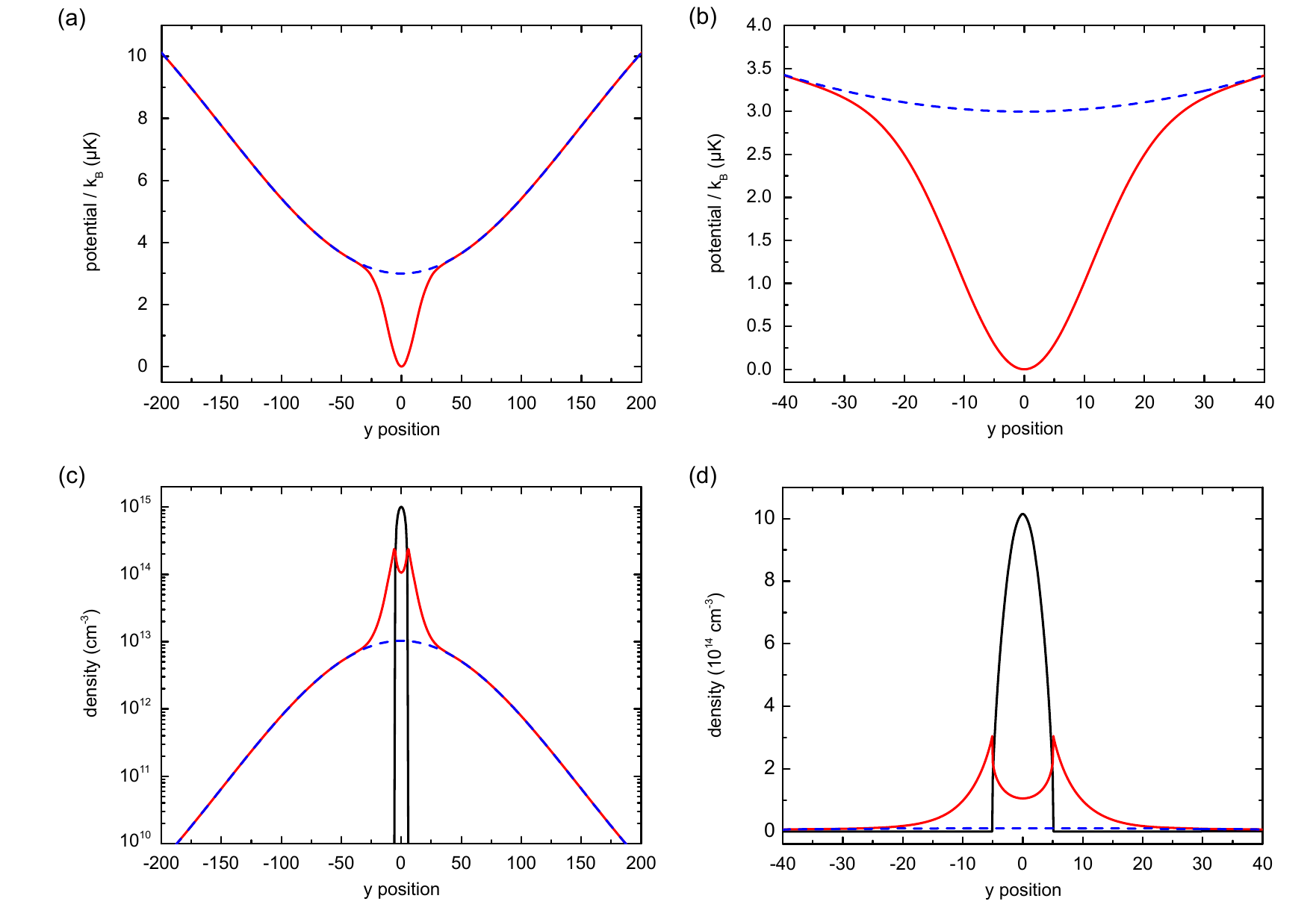}
\caption{\label{fig:Calc:Fig2} Potential (a, b) and density (c, d) along the $y$-direction determined by the numerical model, using parameters that correspond to the star-shaped data point of Fig.~\ref{fig:Calc:Fig1}. In (a) and (b), the reservoir potential is shown as a blue, dashed line and the full potential, including the dimple, as a red line. Panels (c) and (d) show the density distributions of the thermal gas $n_{\rm th}$ (red line) and of the BEC $n_0$ (black line). The dashed blue line shows the density of thermal atoms in absence of the dimple potential $n_{\rm th, res}$.}
\end{figure*}

\subsection{Comparison to the experiment}

We now compare the experiment and the numerical model. Since the model can only describe a thermalized sample without the influence of the laser cooling beams, we perform experiments corresponding to this situation. The experimental sequence is identical to the one described in the Letter, up to the moment before the dimple is ramped on. Here, we switch off laser cooling and we do not use the transparency laser. After ramping on the dimple, we wait 200\,ms to let the system thermalize. Then we record as usual time-of-flight absorption images with 24\,ms expansion time, from which we extract temperature, total atom number and BEC atom number. We perform this measurement for different loading times of the metastable state reservoir at the beginning of the experimental sequence. A longer loading time leads to a higher atom number and, as a consequence of the laser cooling process, to a higher temperature. The result of the experiment is shown in Fig.~\ref{fig:Calc:Fig1}. In the examined range from $3.5\times 10^6$ to $8\times 10^6$ atoms, the temperature changes from $880\,$nK to $990\,$nK. The atom number in the condensate is nearly proportional to the total number of atoms $N_{\rm tot}$ and is well described by $N_0=0.02 N_{\rm tot}$.

\begin{table}[b]
\caption{\label{tab:Calc:Tab1}Parameters of the dipole trap beams used for the experiments described in this section.}
\begin{ruledtabular}
\begin{tabular}{cccccc}
beam  & waist $x$ & waist $y$ & waist $z$ & $P$ & $U/k_B$ \\
&  ($\mu$m) & ($\mu$m) & ($\mu$m) & (mW) & ($\mu$K)  \\
\hline
reservoir & & 298.2 & 17.32 &  1748 & 11.73 \\
vertical  & 297 & 297 & & 743 & 0.29\\
dimple  &  22.41 & 22.41 & & 39.6 & 2.73 \\
\end{tabular}
\end{ruledtabular}
\end{table}

The numerical model needs the potential as input. To describe the potential, we sum the contributions of gravity and of the potentials of the three Gaussian optical dipole trap beams used in this experiment. The parameters of the beams, determined from trap oscillation and beam power measurements, are given in Tab.~\ref{tab:Calc:Tab1}. The precision with which we know the potential shape is not quite high enough to give a good match between the model and the experiment. We therefore vary one potential parameter, here the dimple depth, and match the model to the experiment around $N_{\rm tot}=6\times 10^6$. Experiment and model agree for a dimple depth that is 10\% larger than the measured one. A good match between model and experiment could have alternatively been achieved by varying other parameters or by correcting a possible systematic error in the experimental data. A candidate systematic measurement error, which would be sufficient to bring the numerical model and the experiment into agreement, is an overestimation of the temperature by 5\%. This hypothetical systematic shift is smaller than our experimental uncertainty.

\begin{table}[b]
\caption{\label{tab:Calc:Tab2} Comparison of the experiment and the models. The values given correspond to the star-shaped data point in Fig.~\ref{fig:Calc:Fig1}. The trap oscillation frequencies of dimple and reservoir are $f_x=244\,$Hz, $f_y=247\,$Hz, $f_z=624\,$Hz, $f_{{\rm res},x}=8.4\,$Hz $f_{{\rm res},y}=37\,$Hz, and $f_{{\rm res},z}=622\,$Hz.}
\begin{ruledtabular}
\begin{tabular}{llll}
quantity & experiment & numerical & analytical\\
\hline
T & 944(50)\,nK & 944\,nK & 944\,nK\\
$\mu/k_B$ & & 478\,nK & 405\,nK\\
$N_{\rm BEC}$ & $111\times 10^3$ & $111\times 10^3$& $111\times10^3$\\
$N_{\rm dimple}$ & & $654\times 10^3$ &  \\
$N_{\rm reservoir}$ & & $5.0\times 10^6$ & $2.6\times10^6$\\
$N_{\rm total}$ & $5.8\times10^6$ & $5.8\times 10^6$ & \\
$n_0(0)$ & & $1.0\times 10^{15}\,$cm$^{-3}$ & $8.6\times 10^{14}\,$cm$^{-3}$\\
$n_c$ & & $3.5\times 10^{14}\,$cm$^{-3}$ & $3.5\times 10^{14}\,$cm$^{-3}$\\
$n_{\rm th, edge}$ & & $9.9\times 10^{12}\,$cm$^{-3}$ & $8.7\times 10^{12}\,$cm$^{-3}$\\
$v$ & & $21.8\,\mu$m/ms & $21.8\,\mu$m/ms\\
$\Gamma_{\rm el, th, peak}$ & & 2700\,s$^{-1}$ & 2700\,s$^{-1}$\\
$\Gamma_{\rm el, th, res}$ & & 50\,s$^{-1}$ & 50\,s$^{-1}$\\
$R_{{\rm TF},x}$ & & $5.1\,\mu$m & $5.8\,\mu$m\\
$R_{{\rm TF},y}$ & & $5.0\,\mu$m & $5.8\,\mu$m\\
$R_{{\rm TF},z}$ & & $1.9\,\mu$m & $2.3\,\mu$m\\
\end{tabular}
\end{ruledtabular}
\end{table}

After matching the model to the data in a single point, we can now explore the whole range of data. The model requires as input parameters the chemical potential and the temperature. For two data points near $3.5\times 10^6$ and $8\times 10^6$ atoms, we determine the chemical potential that corresponds to the measured BEC atom number. This procedure results automatically in a good match for the total atom number, which already shows that the model describes the data well. Instead of determining the chemical potential for each measurement point, we now simply linearly interpolate temperature and chemical potential between these two extreme points and calculate $N_0$ and $N_{\rm tot}$ for many values in between. The result is the solid line in Fig.~\ref{fig:Calc:Fig1} and shows a good overall match between experiment and calculation. In a second calculation, we keep the temperature always at $930\,$nK and vary only the chemical potential; see the dashed line in Fig.~\ref{fig:Calc:Fig1}. The mismatch between calculation and data shows the importance of the small temperature change observed in the experiment between low and high atom number.

We now examine the data point marked by a star in Fig.~\ref{fig:Calc:Fig1} in more detail. We chose this data point because the atom number in the BEC is comparable to the one achieved in the experiments discussed in the Letter. As input for the models, we use the measured temperature and BEC atom number and adapt the chemical potential accordingly. Figure~\ref{fig:Calc:Fig2} shows the potential and the density profiles of BEC and thermal gas along the $x$-direction as determined by the numerical calculation. The BEC mean-field clearly has a strong effect on the thermal distribution, pushing thermal atoms outwards. Table~\ref{tab:Calc:Tab2} compares experimental data with the two models and shows a general good agreement.

Some information extracted from the models is worth further discussion. We expect that the conditions in the experiments presented in the Letter are comparable to the experiments presented here within the volume illuminated by the transparency laser. Only outside this volume, laser cooling will change the conditions and make a comparison more complicated. An indication for this is that the atom number in the dimple, which was not measured here, is comparable to the one observed in the experiment of the Letter.

Since the reservoir contains nearly an order of magnitude more atoms than the dimple, the density distribution of the gas in the reservoir and the temperature of the system will barely change when ramping the dimple on. Therefore the ratio of $n_{\rm th, edge}/n_c$, which is 35, gives an idea of the thermal gas density increase and the phase-space density increase provided by the dimple.

Finally, we note that the average elastic collision rate in the reservoir close to the dimple corresponds to about five collisions during the $\sim100\,$ms required for thermalization. This number compares well with the $\sim 3$ collisions required to obtain thermal equilibrium as determined from Monte Carlo simulations.

\bibliographystyle{apsrev}


\end{document}